\documentclass[aps,twocolumn,prb,]{revtex4-2}
\usepackage{amssymb}
\usepackage{amsmath}
\usepackage{bm}
\usepackage{color}
\usepackage{graphicx}
\usepackage{tikz}
\usetikzlibrary{snakes}
\usepackage{hyperref} 

\begin{document}

\title{Heat transport through a superconducting artificial atom}
\author{Meng Xu}
\email{meng.xu@uni-ulm.de}
\author{J. T. Stockburger}
\email{juergen.stockburger@uni-ulm.de}
\author{J. Ankerhold}
\email{joachim.ankerhold@uni-ulm.de}
\affiliation{
Institute  for Complex Quantum Systems and IQST, Ulm University - Albert-Einstein-Allee 11, D-89069  Ulm, Germany}

\date{\today}
\begin{abstract}
Quantum heat transfer through a generic superconducting set-up consisting of a tunable transmon qubit placed between resonators that are termined by thermal reservoirs is explored. Two types of architectures are considered, a sequential and a beam splitter setting.   Applying the numerical exact hierarchical equation
of motion (HEOM) approach, steady state properties are  revealed,  and experimentally relevant
parameter sets are identified. Benchmark results are compared with predictions based on approximate treatments to demonstrate their failure in broad ranges of parameter space. These findings may allow to improve future designs for heat control in superconducting devices.
\end{abstract}

\maketitle

\newpage

\section{Introduction}

Theoretical studies of photonic heat transport
\cite{karimi2016otto,karimi2017coupled,ojanen2008mesoscopic,thomas2019photonic,agarwalla2019photon,moghaddasi2019phononic,yamamoto2020heat} and rectification
\cite{segal05a,ruokola2009thermal,otey2010thermal,motz2018rectification,riera2019dynamically} in superconducting platforms
play an important role not only in understanding current experimental findings \cite{ronzani2018tunable,senior2020heat}, but also in designing and potentially improving future architectures to control heat, for example in circuit Quantum Electrodynamics (cQED) \cite{ronzani2018tunable,senior2020heat,maillet2020electric}. Moreover, fundamental questions regarding signatures of quantum mechanics in thermodynamic properties of devices at nanoscales \cite{binder2019thermodynamics,goold2016role,cui2017perspective} have not been answered yet and require advanced simulation techniques beyond conventional perturbative treatments \cite{hofer2017markovian,mukherjee2020anti,newman2020quantum,vadimov2020validity,majland2020quantum,Aurell2020large}.

Dynamical control of heat flow has received particular attention recently, for example, mediated by phonons in solid state circuits \cite{schwab2000measurement}, in carbon nanotubes   \cite{chang2006solid}, and in arrangements of trapped ions \cite{pruttivarasin2011trapped}. Further, energy transfer due to electronic motion in hybrid systems \cite{martinez2015rectification}, where normal metals are
tunnel-coupled to superconductors, and due to microwave photons in superconducting circuits   \cite{ronzani2018tunable,senior2020heat,meschke2006single,pascal2011circuit} have been studied.

Particularly, superconducting devices as promising platforms for quantum information processing allow to modulate the heat flow carried by photons based on superconducting quantum interference device (SQUID) with tunable magnetic flux, see e.g.\ \cite{ronzani2018tunable}. In the milli-Kelvin (mK) temperature domain, where they are operated, quasi-particle energy transfer is basically absent  \cite{meschke2006single,schmidt2004photon} and residual phononic channels, while not completely avoidable, give, however,  rise to only flux independent energy transfer  \cite{ronzani2018tunable,senior2020heat}. In addition, circuits including Josephson junctions naturally exhibit nonlinear behavior, a pre-requisite for heat rectification \cite{segal05a,motz2018rectification,senior2020heat}.

The theoretical framework to describe heat transfer between two thermal reservoirs at different temperatures in quantum mechanical settings is a challenging task. Mostly, model systems have been explored so far and  often based on approaches, where the interactions between the heat modulating component (system) and the thermal reservoirs are treated perturbatively, for example, in terms of quantum optical master equations \cite{breuer02}. The latter, however, suffer not only from the known limitations to weak couplings and sufficiently elevated temperatures \cite{breuer02,weiss12,breuer16,vega17}, but also from deficiencies related to the proper state representation of the dissipative part in the evolution equation of the reduced density operator \cite{kosloff2014,stockburger2017}.
In contrast, the description of realistic set-ups at cryogenic temperatures has received much less attention \cite{karimi2016otto,karimi2017coupled,
ojanen2008mesoscopic,thomas2019photonic,ronzani2018tunable,senior2020heat}.

The goal of this paper is to fill this gap and to answer, at least partially, the following major questions: First, to which extent are conventional perturbative treatments able to quantitatively describe the modulation of the heat flow through a generic superconducting setting? Second, how does the heat flow depend on circuit properties, either fixed by design or tunable in situ?

For this purpose, in the following we employ the so-called Hierarchical equations of motion (HEOM) formulation
\cite{tanimura89,tanimura06,tanimura2020numerically,kato15,song17a}, a non-perturbative simulation technique of open quantum systems which is derived from the formally exact representation of the reduced density in terms of path integrals \cite{weiss12,feynman63}. The HEOM has the additional benefit that, depending on the depths of the hierarchy, perturbative approaches such as the Redfield  and the Lindblad master equations can easily be computed with the same code. We note in passing that alternative, non-perturbative frameworks such as the
Stochastic Liouville-von Neumann Equation(SLN) \cite{motz2018rectification, stockburger02}, the
multilayer multiconfiguration time-dependent Hartree (ML-MCTDH)
approach\cite{velizhanin2008heat,wang03}, continuous-time quantum Monte Carlo (CT-QMC) algorithm\cite{yamamoto2018heat,gull11} have also been applied recently to explore quantum heat phenomena.

More specifically, we will consider a generic set-up, where a two level system, implemented in form of a transmon qubit, is placed in series between two resonators that are coupled to respective thermal reservoirs, realized as ohmic resistors, see Fig.~\ref{fig:fig1}. Due to the SQUID architecture of the transmon an external magnetic flux allows to tune its transition frequency and thus to modulate the capacity to carry heat through the resonator-transmon-resonator device. This quantum heat valve has been experimentally realized in a slightly more complex circuitry in \cite{ronzani2018tunable}. We consider two scenarios, one in which energy is transferred through the transmon only, and one, in which in addition heat can also directly flow from the left to the right resonator, thus bypassing the transmon. We provide a detailed understanding of how the energy level structure of the total complex (resonator-transmon-resonator) determines the heat flow through  quantum channels in resonant and off-resonant (dispersive) regimes. Benchmark results obtained within the HEOM allow to fix, when compared to experimental data, circuit parameters, particularly those which are not known or not known with sufficient accuracy. In addition information about residual transfer channels (phonons) that are not sensitive to the flux modulation can be retrieved.

The remaining sections of this paper are arranged as follows. In Sec. \ref{sec:model} we present the model, the sequential and the beam-splitter settings, respectively, and their properties in absence of thermal contacts.
 The HEOM approach, how to extract the heat currents, and approximate treatments are briefly discussed in \ref{sec:technique}. The main part of the paper is Sec.~\ref{sec:results}. There, we compare results from the HEOM with those obtained with perturbative treatments, investigate in detail the steady state with constant heat flow from hot to cold reservoir, and  analyze the parameter dependence of the quantum heat valve. This sections also an extension to a quantum heat rectifier and its performance when the symmetry in the setting is broken. Finally, in Sec.\ref{sec:experiment} a detailed comparison with the available data from the experiment that is reported in \cite{ronzani2018tunable} is given. Conclusions are made in Sec. \ref{sec:conclusions}.

\section{Theoretical modelling}
\label{sec:model}
\subsection{Total system Hamiltonian}
\label{subsec:tsh}
\begin{figure}
\centering
\resizebox{8.5cm}{2.04cm}{
\begin{tikzpicture}[scale=1,font=\small]
\node[fill=red!20,draw=black,scale=2] (v0) at (0,0) {$T_L$};
\draw[line width=1] (3,-1) parabola (4,1.5);
\draw[line width=1] (3,-1) parabola (2,1.5);
\draw[fill=green!10] (5.5,0) circle (0.5);
\draw[line width=1] (8,-1) parabola (9,1.5);
\draw[line width=1] (8,-1) parabola (7,1.5);
\node[fill=blue!20,draw=black,scale=2] (v1) at (11,0) {$T_R$};
\draw[-,snake=snake] (0.74,0)--(2.4,0);
\draw[-,snake=snake] (3.6,0)--(5,0);
\draw[-,snake=snake] (6,0)--(7.4,0);
\draw[-,snake=snake] (8.6,0)--(10.26,0);
\draw[-,snake=snake] (3.26,-0.75)--(7.72,-0.75);
\draw[-,line width=1] (2.45,-0.25)--(3.55,-0.25);
\draw[-,line width=1] (2.3,0.25)--(3.7,0.25);
\draw[-,line width=1] (5.06,-0.25)--(5.94,-0.25);
\draw[-,line width=1] (5.06,0.25)--(5.94,0.25);
\draw[-,line width=1] (7.45,-0.25)--(8.55,-0.25);
\draw[-,line width=1] (7.3,0.25)--(8.7,0.25);
\node[scale=1] (s1) at (1.67,-0.3) {$\eta_L$};
\node[scale=1] (s2) at (4.3,0.3) {$g_L$};
\node[scale=1] (s3) at (6.7,0.3) {$g_R$};
\node[scale=1] (s4) at (9.43,-0.3) {$\eta_R$};
\node[scale=1] (s5) at (5.49,-1.3) {$\tilde{g}$};
\node[scale=1] (s6) at (3,0.5) {$\omega_L$};
\node[scale=1] (s7) at (5.5,0) {$\omega_q$};
\node[scale=1] (s8) at (8,0.5) {$\omega_R$};
\end{tikzpicture}
}
\caption{A schematic of the quantum heat valve described in this article.
The total system Hamiltonian is given by Eq. (\ref{Eq:htot}).}
\label{fig:fig1}
\end{figure}
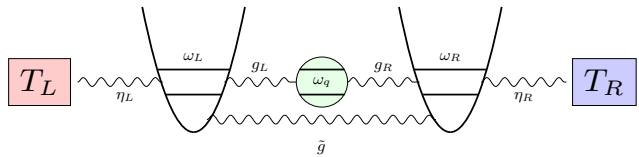

Tunable photonic heat transport and heat rectification can be achieved by
the platform of superconducting circuit quantum electrodynamics (cQED)
\cite{ronzani2018tunable,senior2020heat}. In the language of open quantum
systems \cite{weiss12}, a corresponding Hamiltonian of the total system then can be formally partitioned into three parts
\begin{equation}\label{Eq:htot}
  H = H_s + H_{b} + H_{sb} \;\;,
\end{equation}
where $H_s$ is what one identifies as the system Hamiltonian which can be controlled experimentally to act as a valve \cite{ronzani2018tunable}, $H_b$ describes two thermal reservoirs at different temperatures $T_L\geq T_R$ (henceforth denoted as left bath and right bath, respectively), and $H_{sb}$ is the coupling between these parties. In the sequel, we consider a situation as shown in Fig.~\ref{fig:fig1}, where this coupling to an artificial atom (transmon) is mediated by two harmonic oscillators ($LC$-circuits or cavities). These oscillators can either be included in the system part or in the reservoir part as we will discuss below.

In absence of thermal reservoirs and any internal interactions, the bare Hamiltonian 
of our cQED system reduces to
\begin{equation}\label{Eq:hqs}
  H_0 =
   \hbar\omega_La_L^{\dag}a_L + \hbar\omega_q \sigma_+ \sigma_{-} + \hbar\omega_Ra_{R}^{\dag}a_R + H_{ZPE}\;\; ,
 \end{equation}
where $\hbar\omega_L$, $\hbar\omega_q$ and $\hbar\omega_R$ are the
transition energies of the left resonator, qubit and right resonator, respectively, with standard annihilation and creation operators for the oscillators and Pauli matrices $\sigma_k, k=x, y, z$ with $\sigma_\pm=\sigma_x\pm i \sigma_y$.
The Zero point energy (ZPE) of the bare system is denoted by a constant $H_{ZPE}$, which, however, does not affect the quantum dynamics and will thus be set zero (see also \cite{ronzani2018tunable}).
The frequency of the transmon-qubit $\omega_q$ is tunable via a magnetic flux $\phi$ according to
\begin{equation}\label{}
  \omega_q(\phi) = \frac{\sqrt{8E_J(\phi)E_c} - E_c}{h}
\end{equation}
with charging energy $E_c=2e^2/C$ and effective Josephson energy
\begin{equation}\label{eq:ejphi}
  E_J(\phi) = E_{J0}|\cos(\pi\frac{\phi}{\phi_0})|\sqrt{1+d^2\tan^2(\pi\frac{\phi}{\phi_0})} \;\;,
\end{equation}
where $\phi_0=h/2e$ is the magnetic flux quantum. An asymmetry in the two Josephson junctions constituting the transmon is captured by an asymmetry parameter $d$ \cite{koch07charge}. According to Fig.~\ref{fig:fig1}, internal couplings induce interactions between the transmon and the respective oscillators (couplings $g_L, g_R$) and directly between the oscillators ($\tilde{g}$), respectively.

The bare thermal reservoirs, kept at different temperatures, are modelled as usual by  quasi-continua of independent harmonic degrees of freedom, i.e.,
\begin{equation}\label{}
\begin{split}
  H_b &=  H_L + H_R  \\
      &=
   \sum_{\alpha=L,R} \left\{ \sum_{i=1}^{\infty}
   \frac{p_{\alpha}^2}{2m_{\alpha,i}} + \frac{1}{2}m_{\alpha,i}\omega_{\alpha,i}^2 x_{\alpha,i}^2  \right\}\, ,
\end{split}
\end{equation}
where $m_{\alpha,i}$, $\omega_{\alpha,i}$, $x_{\alpha,i}$ and
$p_{\alpha,i}$ denote mass, frequency, coordinate and
momentum of the $i$th harmonic oscillator in the $\alpha$th reservoir.
The bilinear interaction between these reservoirs and what is considered as the system is given by
\begin{equation}\label{Eq:hsb}
  H_{sb} =  \sum_{\alpha=L, R} X_{\alpha} (a^{\dag}_\alpha + a_{\alpha}) + \frac{1}{2} \sum_{\alpha=L, R} \mu_\alpha (a_{\alpha}^{\dag}+a_\alpha)^2  \;\;,
\end{equation}
with $X_{\alpha} = \sum_{i}c_{\alpha,i}x_{\alpha,i}$ denoting a collective coordinate of the $\alpha$th
reservoir and $c_{\alpha,i}$ being the coupling constant of the $i${\rm th} mode with oscillator $\alpha$.
The last term guarantees that the reservoirs only act dynamically upon the system and any coupling induced distortion is absent (counter term).  In the following, we will put masses to unity and $\hbar = 1$.

In this setting the effective impact of the reservoirs onto the system is completely described by coupling weighted spectral densities
\begin{equation}
\label{Eq:dsp}
  J_{\alpha}(\omega) = \frac{\pi}{2}\sum_{i}
  \frac{c_{\alpha,i}^2}{\omega_{\alpha,i}}\delta(\omega - \omega_{\alpha,i}) \;\;
\end{equation}
which implies $\mu_\alpha=(2/\pi) \int_0^\infty d\omega J_\alpha(\omega)/\omega$. In the following, we assume a generic ohmic spectral density with a Debye type of frequency cut-off $\omega_{{\rm D}, \alpha}$, i.e.,
\begin{equation}\label{eq:debye}
  J_{{\rm D}, \alpha}(\omega) = \frac{\eta_{\alpha}\,  \omega}{1+\omega^2/\omega_{{\rm D}, \alpha}^2} \;\;
\end{equation}
with a dimensionless coupling $\eta_\alpha$. This way, an unmanageable UV divergence of purely Ohmic dissipation is avoided.

\subsection{Sequential setting}
\label{subsec:seq}
As a first case, we assume the resonators and the qubit to be coupled in a linear geometry such that no direct resonator-resonator interaction exists, i.e.\ $\tilde{g}=0$ in Fig.~\ref{fig:fig1}. As a consequence,  this sequential setting is described by $H_s\equiv H_{\rm seq}$ with
\begin{equation}\label{Eq:shsys}
H_{\rm seq}=H_0+\sum_{\alpha=L,R} g_{\alpha}(a_{\alpha}^\dagger+
  a_{\alpha})\, (\sigma_+ +\sigma_{-})\, .
\end{equation}

However, in a sequential setting one can also consider the intermediate resonators as part of the respective reservoirs to which they are directly coupled. Technically, this is achieved by performing a  normal mode transformation \cite{garg1985effect} of the so-defined total
Hamiltonian and eventually leads to a spin-boson form for the total Hamiltonian \cite{segal05a,thoss2001self} with $H_s$ representing the transmon only, i.e.,
\begin{equation}\label{Eq:effecth}
\begin{split}
  H =& \omega_q \sigma_+ \sigma_{-} + (\sigma_+ +\sigma_{-})\sum_{\alpha,i}^{}
  \tilde{c}_{\alpha,i}\tilde{x}_{\alpha,i} \\
  &+ \sum_{\alpha,i}^{} \left\{
  \frac{\tilde{p}_{\alpha,i}^{2}}{2} + \frac{1}{2}
  \tilde{\omega}_{\alpha,i}^2\tilde{x}_{\alpha,i}^2
  \right\}\;\;
\end{split}
\end{equation}
up to an irrelevant constant, where quantities with a tilde denote those of the mapped reservoirs.

For sufficiently large cut-off frequency, the spectral distribution (\ref{Eq:dsp}) transforms accordingly such that it
takes a Lorentzian form in the spin-boson representation \cite{garg1985effect,thoss2001self,iles2014environmental}
\begin{equation}\label{Eq:jeff}
  J_{{\rm eff}, \alpha}(\omega) =\kappa_{\alpha}\,  \frac{{\tilde{\eta}}_{\alpha}\omega}{(1-\omega^2/\omega_{\alpha}^2)^2 + {\tilde{\eta}}_{\alpha}^2\omega^2/\omega_\alpha^4 } \;\;
\end{equation}
with $\kappa_{\alpha} = 2g_{\alpha}^2/\omega_{\alpha}^3$.  In the low frequency sector $\omega\ll \omega_\alpha$ the ohmic-type of dissipation is recovered with $J_{{\rm eff},\alpha}\approx \kappa_\alpha \tilde{\eta}_\alpha\omega $ and effective coupling $\kappa_\alpha \tilde{\eta}_\alpha$.
We note in passing that $\tilde{\eta}_{\alpha}$ has the dimension of a frequency in contrast to the dimensionless $\eta_{\alpha}$ in Eq. (\ref{eq:debye}); since $\kappa_{\alpha} \tilde{\eta}_{\alpha}$ is dimensionless, both $J_{\rm eff}$ and $J_{\rm D, \alpha}$ carry identical dimensions.

This transformation reduces the system Hilbert space considerably,
 (from resonator-qubit-resonator to qubit) and it leads to an effective spectral distribution through which the qubit interacts with effective reservoirs, where the intermediate oscillators appear as frequency filters. While for conventional perturbative treatments of open system dynamics in terms of master equations, this may cause substantial problems with all interesting situations  $\omega_\alpha\approx \omega_q$ being outside the range of validity, it implies a profound increase in efficiency for the non-perturbative HEOM that we apply here.

This can also been seen upon inspection of the auto-correlation functions of the transformed $\alpha$-th reservoir correlation function $\tilde{C}_\alpha(t)=\langle \tilde{X}_\alpha(t)\tilde{X}_\alpha(0)\rangle_\beta$ that are obtained as \cite{tanimura94,tanaka09,ikeda2020generalization}
\begin{equation}\label{Eq:bwct}
\begin{split}
  \tilde{C}_{\alpha}(t) =& \frac{1}{\pi} \int_{-\infty}^{+\infty}d\omega J_{{\rm eff}, \alpha}(\omega) \frac{e^{-i\omega t}}{1-e^{-\beta_{\alpha}\omega}} \\
 =&\frac{\kappa_{\alpha}\omega_{\alpha}^4}{4\xi_{\alpha}}
\sum_{\sigma=\pm 1} \left\{\coth[\frac{\beta_{\alpha}}{2}
  (\xi_{\alpha}-i\sigma \frac{{\eta}_{\alpha}}{2})] + \sigma\right\} \\
  & \times
  \exp\left[-(\frac{{\eta}_{\alpha}}{2}+i \sigma\xi_{\alpha})t\right] \\
-&\frac{2\kappa_{\alpha}\eta_{\alpha}\omega_{\alpha}^4}{\beta_{\alpha}}
\sum_{k=1}^{\infty} \frac{\nu_k}{(\omega_{\alpha}^2 + \nu_k^2)^2 -{\eta}_{\alpha}^2\nu_k^2} e^{-\nu_k t} \;\;.
\end{split}
\end{equation}
where $\xi_{\alpha} = \sqrt{\omega_{\alpha}^2 - {\eta}_{\alpha}^2/4}$,
  $\nu_k = 2\pi k/\beta_{\alpha}$ are the Matsubara frequencies with inverse temperature $\beta_{\alpha}=1/(k_BT_{\alpha})$. Apparently, the correlation functions decay to zero with rates $\eta_\alpha$ in the underdamped limit and at higher temperatures, while at low temperatures the last term including Matsubara frequencies becomes very long-ranged on the order $\hbar\beta$, thus inducing strong retardation effects. We note in passing that the quantum dynamics of spin-boson type of  systems have been widely studied in the past \cite{segal05a,motz2018rectification,velizhanin2008heat, thoss2001self,iles2014environmental,segal14,leggett87,tamascelli2018nonperturbative,duan17} employing both perturbative as well as non-perturbative techniques.

\subsection{Beam-splitter setting}
\label{subsec:obof}
For the second case, we extend the sequential setting by allowing also a direct resonator-resonator interaction with strength $\tilde{g}$. Accordingly, energy transfer can happen via two channels, one which includes the qubit and one which does not (beam splitter setting).
The system part of the total Hamiltonian thus takes the form $H_s\equiv H_{\rm beam}$ with
\begin{equation}\label{Eq:bhsys}
\begin{split}
H_{\rm beam} =& H_0 + \sum_{\alpha=L,R}
g_{\alpha}(a_{\alpha}^{\dag}+ a_{\alpha})(\sigma_++\sigma_{-}) \\
&+ \tilde{g}(a_{L}^{\dag}+a_{L})(a_{R}^{\dag}+a_{R}) \;\;
\end{split}
\end{equation}
with the coupling to the reservoirs again governed by $q_\alpha$.
This beam splitter situation does not allow for a reduction similar to the sequential setting: If the resonators are incorporated in the respective reservoirs, an effective reservoir-reservoir interaction emerges which does not provide any benefit compared to the original model.

The spectral density (\ref{Eq:dsp}) gives then rise to
the auto-correlation function of the $\alpha$th bath $C_\alpha(t)=\langle X_\alpha(t) X_\alpha(0)\rangle_\beta$ as
\begin{equation}\label{Eq:dbct}
\begin{split}
  C_{\alpha}(t) =& \frac{1}{\pi}\int_{-\infty}^{+\infty} d\omega J_{{\rm D}, \alpha}(\omega) \frac{e^{-i\omega t}}{1-e^{-\beta_{\alpha} \omega}} \\
  =& \frac{\eta_{\alpha}\omega_{{\rm D}, \alpha}^2}{2}[\cot(\frac{\beta_{\alpha}\omega_{{\rm D}, \alpha}}{2})-i] e^{-\omega_{{\rm D},\alpha} t}\\
  &+\frac{2\eta_{\alpha}\omega_{{\rm D}, \alpha}^2}{\beta_\alpha} \sum_{k=1}^{+\infty} \frac{\nu_k}{\nu_k^2-\omega_{{\rm D}, \alpha}^2}\, e^{-\nu_k t} \;\;.
\end{split}
\end{equation}
As we will see, the beam-splitter interaction makes the quantum dynamics richer compared to that of the sequential setting. So far, it has  been studied only in limiting regimes using perturbative methods \cite{ronzani2018tunable,reuther2010two}.

\subsection{Bare system energy spectrum}
\label{subsec:sec-bsp}

\begin{figure}[htbp]
\centering
\includegraphics[width=8cm]{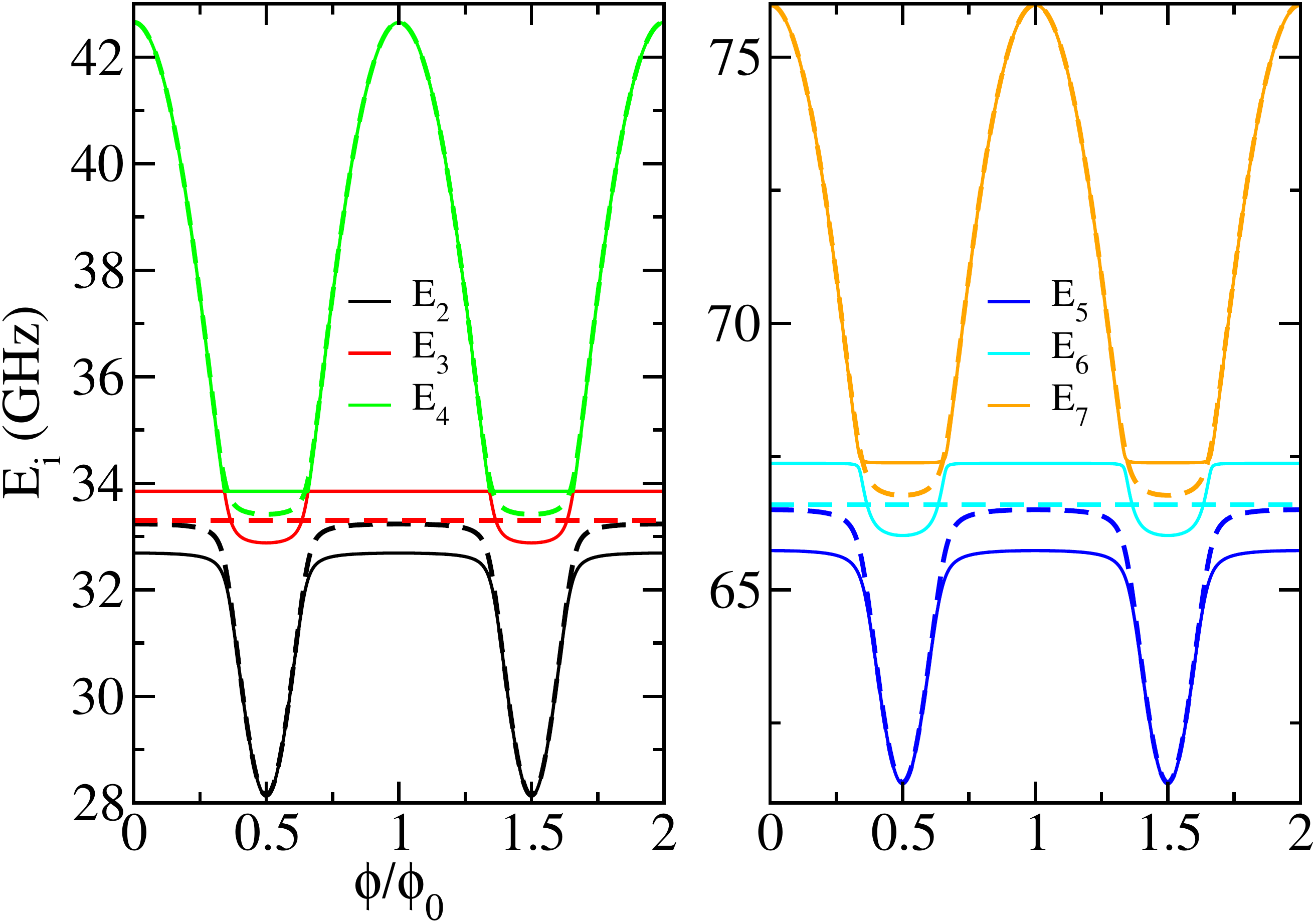}
\caption{Energy spectra of bare system Hamiltonians (sequential, beam splitter) in Eqs. (\ref{Eq:shsys}) and (\ref{Eq:bhsys}). Here, the relevant states are represented as: $|1\rangle = |000\rangle$, $|2\rangle = |100\rangle$, $|3\rangle = |010\rangle$,
$|4\rangle = |001\rangle$, $|5\rangle = |110\rangle$, $|6\rangle = |101\rangle$ and $|7\rangle = |200\rangle$ with corresponding eigenenergies $E_1,...,E_9$, where we put $E_1=0$.
Solid lines denote the beam-splitter, dashed lines the
sequential setting. Parameters are:
$E_{j0}/2\pi = 40$ GHz, $E_c/2\pi = 0.15$ GHz, $d = 0.45$, $g_L = g_R = 0.55$ GHz, $\tilde{g} = -0.55$ GHz, $\hbar\omega_L = \hbar\omega_R = 33.3$ GHz.}
\label{fig:fig2}
\end{figure}

Before we start with a discussion of the heat flow through the respective settings, it is instructive to study the energy spectra of the bare systems, see Fig.~\ref{fig:fig2}. Here, we use parameters typical for superconducting settings (see e.g.~\cite{ronzani2018tunable}).
As one expects, the impact of the direct oscillator-oscillator coupling is most prominent far away from resonances between oscillators and transmon, i.e. in the so-called dispersive regime $|\Delta_\alpha|\gg g_{\alpha}$ with de-tuning $ \Delta_\alpha = \omega_q-\omega_{\alpha}$.
Indeed, in this regime and for symmetric couplings $g_{\alpha}=g$, $\Delta_{\alpha}=\Delta$,  the Hamiltonian $H_{\rm beam}$ in (\ref{Eq:bhsys}) can be cast via a Schrieffer-Wolf transform \cite{schrieffer1966relation,haq2020systematic} and retaining only terms up to second order in $(g/\Delta)^2$ \cite{reuther2010two,mariantoni2008two,ciani2017three}  into the approximate form
\begin{equation}\label{Eq:heff}
\begin{split}
& H_{\rm beam}^{\rm eff} = e^{S}H_{\rm beam}e^{-S} \\
\approx &\; (\omega_q + \frac{2g^2}{\Delta})\sigma_+ \sigma_{-} +
(\omega_L + \frac{g^2}{\Delta}\sigma_z)a_L^{\dag}a_L\\
+&(\omega_R + \frac{g^2}{\Delta}\sigma_z)a_R^{\dag}a_R
+ (\tilde{g}+ \frac{g^2}{\Delta}\sigma_z)
(a_L^{\dag}a_R + a_La_R^{\dag})\;,
\end{split}
\end{equation}
with
\begin{equation}\label{Eq:sw}
  S = \frac{g}{\Delta}(\sigma_+ a_L -\sigma_{-}a_L^{\dag}
  +\sigma_{+}a_R -\sigma_{-}a_R^{\dag}) \;\;.
\end{equation}

\begin{figure}[htbp]
\centering
\includegraphics[width=8cm]{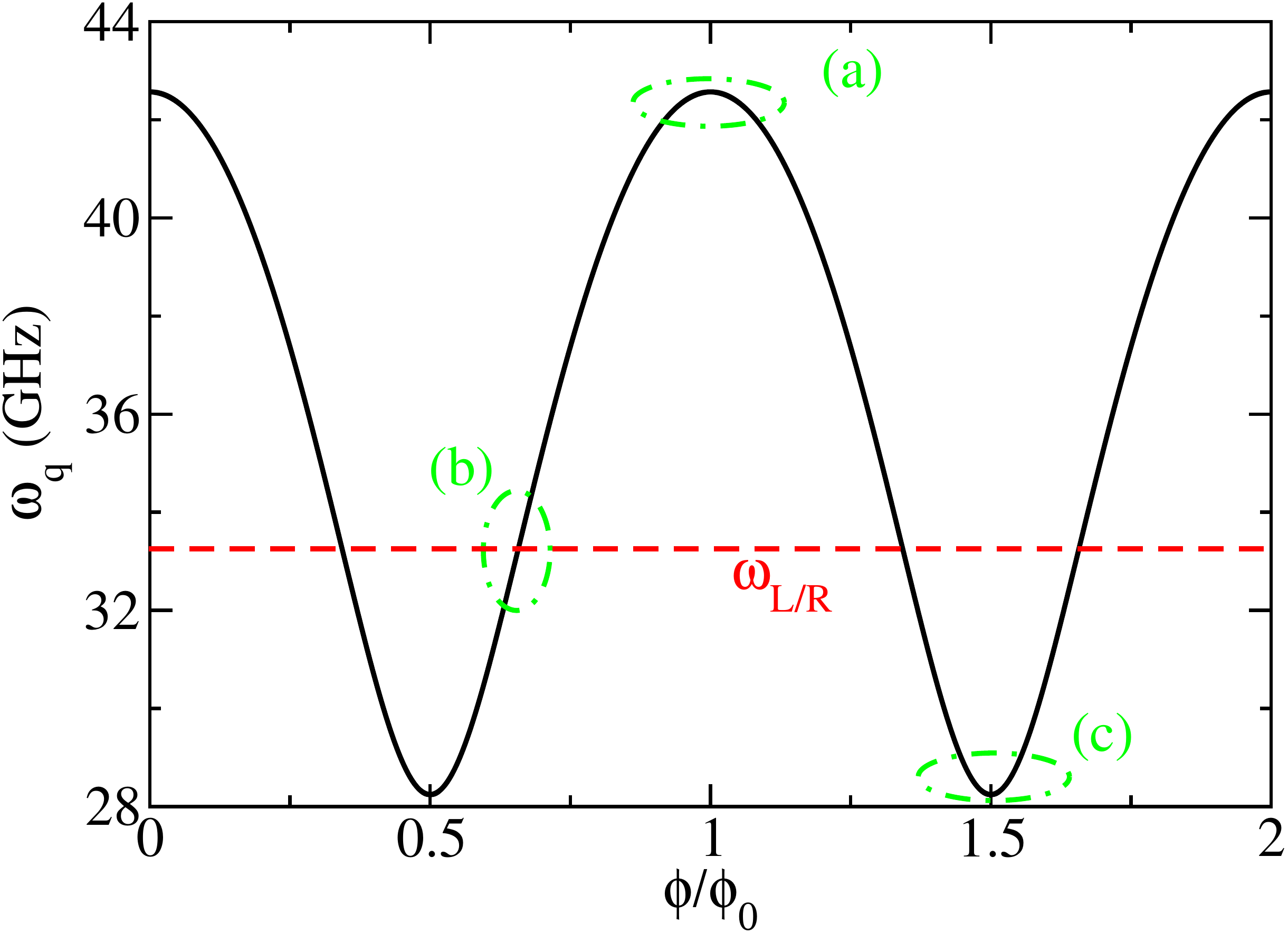}
\caption{Frequency of the transmon-qubit tuned by an external magnetic flux. The parameters are: $E_{j0}/2\pi = 40$ GHz, $E_c/2\pi = 0.15$ GHz and $d = 0.45$. The black line denotes the transmon-qubit frequency, the red line  resonator frequencies.}
\label{fig:fig3}
\end{figure}

Apparently, the oscillator-transmon coupling appears only as renormalization to the oscillator-oscillator coupling. For the parameters chosen in Fig.~\ref{fig:fig2} with equal bare oscillator frequencies $\omega_L=\omega_R$, the dispersive regime with well separated levels between oscillators and transmon is seen in the regimes around $\phi/\phi_0= k$ with $k$~integer (see $E_2, E_3, E_5, E_6$) as well as around $\phi/\phi_0=k+\frac{1}{2}$ with $k$ integer (see $E_4, E_7$), cf.\ also domains (a) and (c) in  Fig.~\ref{fig:fig3}. For finite oscillator-oscillator coupling the degeneracy is lifted and delocalized superpositions of bare oscillator eigenstates appear. For values of the magnetic flux slightly below and above $\phi/\phi_0=k+\frac{1}{2}$ ($k$ integer) avoided crossings indicate resonances between the oscillators and the transmon, see domain (b) in Fig.~\ref{fig:fig3}.

Now, based on this analysis, we expect for the energy transfer across the structure the following behavior: In case of the sequential situation, the energy transferred from hot to cold reservoir approaches maxima at and near  transmon-oscillator resonances $\omega_q(\phi)\approx \omega_L=\omega_R$ \cite{segal05a,may11} and is small otherwise \cite{reuther2010two}. The beam-splitter displays these two maxima as well but in addition exhibits substantial transmission also in regimes, where the direct oscillator channel prevails. More specifically, in the latter regimes an effective
oscillator-oscillator coupling $\tilde{g}_{\rm eff}= \tilde{g}+(g^2/\Delta) \langle \sigma_z\rangle$ has for $\tilde{g}<0$ and at low temperatures a larger absolute value around $\phi/\phi_0= k$ compared to that around $\phi/\phi_0= k+\frac{1}{2}$ due to $\langle\sigma_z\rangle<0$.

For a quantitative description, one could rely on perturbative approaches such as quantum optical master equations. However, it is worth  to discuss a subtlety here that may easily be overlooked. Naively, one may assume that such a treatment is justified as long as $\omega_L, \omega_R, \omega_q\gg \eta_L, \eta_R$ and temperatures are sufficiently elevated. However, these master equations are derived using second order perturbation theory within the eigenstate basis of the system Hamiltonian $H_s$. The more precise and also much stricter condition for their validity is thus $|E_i-E_j|/\hbar\gg \eta_L, \eta_R$ with $E_i, E_j$ being eigen-energies of $H_{\rm seq}$ and $H_{\rm beam}$, respectively. Since avoided crossings lead to energy splittings on the order of the intra-system couplings,
 standard master equations may fail whenever thermal couplings constants $\eta_L, \eta_R$ are on the order of  $\tilde{g}_\alpha, g_\alpha$ which in turn limits their applicability \cite{vadimov2020validity}. For this reason, in the following in order to provide benchmark data, a non-perturbative treatment is employed, the so-called Hierarchical equation of motion (HEOM).

\section{Simulation Techniques}\label{sec:technique}
\subsection{Hierarchical equation of motion}
\label{sec:heom}
Here, we briefly describe the essence of the HEOM approach and how it is derived from the path integral expression of the reduced density matrix. For the sake of simplicity, we consider only a single reservoir for a Hamiltonian of the form (\ref{Eq:htot}) and assume
factorized initial states at time zero,
$\rho_T(0) = \rho_s(0)\otimes e^{-\beta H_b}/Z_b$, where $Z_b = \rm{Tr}\, e^{-\beta H_b}$. The generalization to correlated initial states has also been discussed \cite{tanimura14,song15b}.

Accordingly, in path integral representation \cite{feynman63} the reduced density operator  is obtained as
\begin{equation}\label{Eq:rdo}
\begin{split}
  \rho_s(t)=&\int\mathcal{D}q^{+}(t)\mathcal{D}q^{-}(t) e^{i\{S_{+}[q^{+}(t)]-S_{-}[q^{-}(t)]\}}\\
  & \times\mathcal{F}[q^{+}(t),q^{-}(t)]\rho_s(0) \;\; ,
  \end{split}
\end{equation}
where $q^{+}(t)$ and $q^{-}(t)$ denote forward and backward system paths, respectively, and
$S_{+}[q^+(t)]$ and $S_{-}[q^{-}(t)]$ the corresponding
actions. 
Generally, a continuous system coordinate can be discretized using a system-specific discrete variable representation (DVR) \cite{echave92}. Specially, within the HEOM and for harmonic oscillators and qubits considered here, an eigen-state representation is convenient.
The effective impact of the reservoir onto the system dynamics is described by the influence functional \cite{feynman63} which reads
\begin{equation}\label{Eq:fv-if}
\begin{split}
  &\mathcal{F}[q^{+}(t),q^{-}(t)] =
   \exp\left\{ -\int_{0}^{t}ds [q^{+}(s)-q^{-}(s)] \right. \\
&\times \left.\int_{0}^{s}d\tau
  \left[C(s-\tau)q^{+}(\tau) - C^{\ast}(s-\tau)q^{-}(\tau) \right]
  \right\}\;\;.
\end{split}
\end{equation}

The derivation of the real-time HEOM starts by first
expanding the bath correlation function as a sum of exponential terms Refs.~\cite{tanimura89,jin08}, i.e.
\begin{equation}\label{Eq:cdcp}
\begin{split}
  C(t) &= \frac{1}{Z_b}{\rm Tr}\left[e^{-\beta H_b} X(t)X(0) \right] \\
  &= \frac{1}{\pi}\int_{-\infty}^{+\infty}d\omega J(\omega)\, n_\beta(\omega)e^{i\omega t} \\ &
  =\sum_{k} d_k e^{-\gamma_k t}\;\; \rm{for}\; t>0\;\;
\end{split}
\end{equation}
with the collective bath operator $X$ as in (\ref{Eq:hsb}) and the Bose-Einstein distribution $n_\beta(\omega)$. The $d_k$ denote proper coefficients in an expansion in terms of exponentials with proper rates $\gamma_k$. By the following auxiliary density operator definition,
\begin{subequations}\label{eq:ado1}
\begin{equation}
\begin{split}
\rho_{\bf n}(t)
=&\int \mathcal{D}q^+(t)\mathcal{D}q^-(t)
e^{i\{S_{+}[q^+(t)]-S_{-}[q^-(t)]\}}  \\
&\times \prod_{k} \; [\phi_k(t)]^{n_k}\;
\mathcal{F}[q^+(t),q^-(t)] \; \rho_s(0)\;\;;
\end{split}
\end{equation}
\begin{equation}
\begin{split}
\phi_k(t) =  -i\int_0^t dt_2
&\Big[q^+(t_2)d_k e^{-\gamma_k(t-t_2)}  \\
& -q^-(t_2) d_k^{*} e^{-\gamma_k(t-t_2)} \Big]\;\;,
\end{split}
\end{equation}
\end{subequations}
the formulation of the HEOM leads to \cite{tanimura89,ishizaki05,tanimura06,xu17,zhang2020hierarchical,ikeda2020generalization,yan2020new,tanimura2020numerically}
\begin{align}
\label{eq:real}
\frac{d\rho_{\bm n}(t)}{dt}=&-\left(i\mathcal{L}_s
+\sum_{k}n_{k}\gamma_{k}\right){\rho}_{\bf{n}}(t)-
i\left[\hat{q},\sum_{k}
{\rho}_{{\bf{n}}_{k}^{+}}(t)\right]\nonumber\\
&-i\sum_{k}n_{k}\left(d_{k} \hat{q}
{\rho}_{{\bf{n}}_{k}^{-}}(t)-d_{k}^{*}{\rho}_{{\bf{n}}_{k}^{-}}(t)\hat{q}\right) \;\;,
\end{align}
where $\hat{q}$ denotes the system's coordinate operator that interacts with the collective bath force.
The
$\rho_{\bf{n}}$s are auxiliary density operators
(ADOs) with the subscript $\bf{n}$ denoting a set of integers
$\{n_{1},...,n_{k},...\}$, with $n_{k} \geq 0$ associated with
the $k$th exponential term in Eq. (\ref{Eq:cdcp}); ${\bf{n}}_{k}^{+}$ and ${\bf{n}}_{k}^{-}$ denote
$\{n_{1},...,n_{k}+1,...\}$, and $\{n_{1},...,n_{k}-1,...\}$, respectively.
Further, the super-operator acting on these densities is defined as $\mathcal{L}_s{\rho_{\bm{n}}}=\left[H_{s}, \rho_{\bm{n}}\right]$
with the reduced density operator $\rho_{s}=\rho_{\{0,...0,...\}}$.

Assuming that $\rho_{\bm 0}(t)$ is of order one, the magnitude of
$\rho_{\bm n}(t)$ is proportional to $\prod_k d_k^{n_k}\sim |C(t=0)|^{n_1+...+n_k+...}$, which may be divergent for strong system-bath coupling $\eta$ as $|{\bm n}|\doteq n_1 + n_2 + ....+n_k+...$ increases. Therefore, the unscaled original HEOM \cite{tanimura89,tanimura90} is of not good use in practice. The proposed scaled HEOM combined with on-the-fly filtering methods \cite{shi09b} solves this problem efficiently.
In our simulations, we choose the following rescaling,
\begin{equation}
    \tilde{\rho}_{\bm n}(t) =
    \left(\prod_k n_k! \; |C(0)|^{n_k}\right)^{-1/2}\rho_{\bm n}(t)\;\;,
\end{equation}
so that Eq. (\ref{eq:real}) is recast as
\begin{small}
\begin{equation}\label{eq:scaled}
\begin{split}
\frac{d \tilde{\rho}_{\bf{n}}(t)}{dt}=&-\left(i\mathcal{L}_s
+\sum_{k}n_{k}\gamma_{k}\right) \tilde{\rho}_{\bf{n}}(t)  \\
&-i\sum_k \sqrt{(n_k+1)|C(0)|}\left[\hat{q},
\tilde{\rho}_{{\bf{n}}_{k}^{+}}(t)\right] \\
&-i\sum_{k}\sqrt{\frac{n_k}{|C(0)|}}\left(d_{k} \hat{q}
\tilde{\rho}_{{\bf{n}}_{k}^{-}}(t)-d_{k}^{*}\tilde{\rho}_{{\bf{n}}_{k}^{-}}(t)\hat{q}\right) \;\;,
\end{split}
\end{equation}
\end{small}
where $C(0)=C(t=0)$. The magnitude of $\tilde{\rho}_{\bm n}(t)$ is proportional to $\prod_k \sqrt{|C(0)|^{n_k}/n_k!}$ and decays to zero
for high hierarchical levels. Therefore, we can put $\rho_{\bm n}(t) = 0$ if $|\rho_{\bm n}^{{\rm max}}(t)|<\delta$, where $\delta$ denotes the error tolerance ( here we set $\delta=10^{-7}$). More advanced algorithms to support the efficiency and numerical stability can be found in refers \cite{cui2019highly,shi2018efficient,dunn2019removing,yan2020new,ikeda2020generalization,zhang2020hierarchical,tanimura2020numerically}.

\subsection{Perturbative treatment}
\label{subsec:sec-some}

Approximate treatments of open system dynamics have been developed in second order perturbation theory in the system-reservoir coupling. Together with a time scale separation between fast decaying reservoir correlations and relaxation dynamics of the reduced density operator, this leads to the Redfield master equation. Interestingly, an extended Redfield equation is also obtained from the HEOM if it is cutoff to  $1^{st}$-order ADOs ($\sum_kn_k = 1$). Namely, in the interaction picture this leads to
\begin{subequations}
\begin{equation}\label{Eq:drhos}
  \frac{d}{dt}\rho_s^{I}(t)
  = -i\sum_k[\hat{q}^{I}(t),\rho_{{\bf 0}_k^{+}}^{I}(t)] \;\;,
\end{equation}
\begin{equation}\label{Eq:drho1}
\frac{d}{dt}\rho_{{\bf 0}_k^{+}}^{I}(t)
= -\gamma_k\rho_{{\bf 0}_k^{+}}^{I}(t)
 -i[d_k\hat{q}^{I}(t)\rho_s^I(t) - d_k^{*}\rho_s^I(t)\hat{q}^{I}(t)] \;\;,
\end{equation}
\end{subequations}
where $\hat{q}^I(t)$, $\rho_s^{I}(t)$ and $\rho_{\bm{0}_k^{+}}^{I}(t)$ denote
system coordinate, reduced density matrix and $1^{st}$-order ADOs in the interaction picture, respectively.
Solving Eq. (\ref{Eq:drho1}) and inserting it  into Eq. (\ref{Eq:drhos}) gives
\begin{equation}
\label{Eq:heom2rd}
\begin{split}
\frac{d}{dt}\rho^{I}_{s}(t)
=& -\sum_{k}\int_0^{t}d\tau
e^{-\gamma_k(t-\tau)}
{\bf[} \hat{q}^I(t),
d_k\hat{q}^I(\tau)\rho_s^I(\tau)\\
&- d_k^{*}\rho_s^{I}(\tau)\hat{q}^I(\tau){\bf]}  \\
=&-\int_{0}^{t}d\tau
[q^{I}(t),C(t-\tau)\hat{q}^{I}(\tau)\rho_s^{I}(\tau)\\
&-C^{*}(t-\tau)\rho_s^I(\tau)\hat{q}^I(\tau)] \\
=&-\int_0^t d\tau {\rm Tr_B}\{
[H_{sb}^I(t),[H_{sb}^I(\tau),\rho_T^I(\tau)]]\}\;\;.
\end{split}
\end{equation}
In the above expression, the correlation function in Eq. (\ref{Eq:cdcp}) and the Born approximation  \cite{breuer02} have been used but {\em not} the Markov approximation. Thus,  equation (\ref{Eq:heom2rd}) is a non-local in time integro-differential equation which we call Redfield-plus in the sequel since it goes beyond the conventional Redfield formulation in that the reduced density appears on the right hand side at the instantaneous time $\tau$ and not the final time $t$. This evolution equation can be
solved with the help of auxiliary variables \cite{frishman96,meier99,thanopulos08}.
An additional Markov approximation where we put  $\rho_s^I(\tau)\approx \rho_s^I(t)$, leads to the time-local Redfield master equation
\begin{equation}\label{Eq:redfield}
\frac{d}{dt}\rho^{I}_{s}(t)
=-\int_0^t d\tau {\rm Tr_B}\{
[H_{sb}^I(t),[H_{sb}^I(\tau),\rho_T^I(t)]]\}\;\;.
\end{equation}
Hence, the HEOM approach can been seen as
an infinity-order extension beyond the Redfield-plus/Redfield approximation \cite{xu17,xu2018convergence,trushechkin2019higher}. This
allows to reveal consistently the impact of higher order system-reservoir correlations which are particularly subtle for heat currents.

\subsection{Heat current}
In the framework of the HEOM, effects of the environment onto the system dynamics and statistical properties of the environment can be obtained from
the ADOs \cite{song17a,zhu12,kato15,kato2016quantum,duan2020unusual},. Here, we concentrate on the quantum heat current which is linear in the collective bath force $X$.
One starts, in the interaction picture, with the following two equations
\begin{subequations}
\begin{equation}
  \frac{d}{d t}\rho_s^I(t) = -i\sum_{k}[\hat{q}^I(t),\rho_{\bm{0}_{k}^{+}}^{I}(t)] \;\;;
\end{equation}
\begin{equation}
\begin{split}
\frac{d}{d t}\rho_s^I(t)
= &-i{\rm Tr}_{b}\{[\hat{q}^I(t)X^I(t),\rho_T^{I}(t)] \}\\
=& -i[\hat{q}^{I}(t),{\rm Tr}_{b}\{X^I(t)\rho_T^I(t)\}] \;\; ,
\end{split}
\end{equation}
\end{subequations}
where $\rho_{T}^{I}(t)$ denotes
total density matrix in the interaction picture. In the next step,  relations between $1^{st}$-order ADOs ($\rho_{\bm{0}_{k}^{+}}$)
in the HEOM and $1^{st}$-order moments of $X$ are constructed according to
\begin{equation}\label{}
  \sum_{k}\rho_{\bm{0}_k^{+}}^{I}(t) = \rm{Tr}_{b}\{X^I(t)\rho_T^I(t)\}\;\;.
\end{equation}
Higher-order relations can be found in Refs. \cite{song17a,zhu12}.

In presence of two thermal reservoirs, it is the above relation which is inserted into the definition
\cite{song17a,velizhanin2008heat,wang2014nonequilibrium} of
 the quantum heat current $I_{\alpha}(t)$ between quantum system and the $\alpha$th bath, i.e.
\begin{equation}\label{Eq:qhc}
\begin{split}
I_{\alpha}(t)
  &\equiv -\frac{d}{dt}\langle H_{\alpha}+H_{{\rm sb}, \alpha}\rangle
 = -i\langle [ H_{s},\hat{q}X_{\alpha}]\rangle \\
 &=-i{\rm Tr}_s\left\{[H_s,\hat{q}]
 {\rm Tr}_{b}\{X_{\alpha}\rho_T\} \right\} \\
 &=-i\sum_{k}{\rm Tr}_s\left\{[H_s,\hat{q}]\rho_{\bf{0}_{k}^{+}}(t)\right\} \;\;.
\end{split}
\end{equation}
Here, $H_{{\rm sb}, \alpha}$ denotes the coupling between system and the $\alpha$-the reservoir. Eventually, the total heat current from left (hot) to right (cold) bath is given as
\begin{equation}\label{Eq:hcurent}
    I(t) = \frac{1}{2}[I_{L}(t)-I_{R}(t)] \;\;.
\end{equation}
With the above set of equations at hand, we are in a position to explore heat transfer properties of the two respective settings in more detail.

\section{Results}
\label{sec:results}

In this section, we set the initial factorized state 
$\rho_T(0) = \rho_s(0)\otimes \prod_{\alpha} e^{-\beta_{\alpha} H_{\alpha}}/{\rm Tr}[e^{-\beta_{\alpha}H_{\alpha}}]$ in the simulations. Results for heat transport according to the sequential and the beam-splitter setting, respectively, are presented and discussed. First, we compare data obtained from the approximate  treatments with exact ones from the HEOM. For this purpose, a symmetric situation is considered $\omega_L=\omega_R$ and $g_L=g_R$ which in a recent experiment has been shown to realize a heat valve \cite{ronzani2018tunable}.
Second, for the same scenario and using the HEOM steady state and heat transfer properties are explored for broad ranges of relevant parameters. Third, by breaking symmetry via $\omega_L\neq \omega_R$ we address the situation, where the heat valve turns into a heat rectifier as implemented in \cite{senior2020heat}.

In the HEOM numerical simulations, P\'ade decompositions\cite{hu10,hu11} and on-the-fly
filtering \cite{shi09b} are adopted in order to achieve high efficiency.
Further, Gigahertz (GHz) units are applied in the sequel, typical for superconducting circuits.
Since we focus on the low temperature quantum domain $\beta_{L/R}\hbar\omega_{L/R}> 1$, convergence is reached by working in a restricted Hilbert space spanned by a basis of seven eigenstates $|n_L n_q n_R\rangle$ of the bare oscillators ($|n_L\rangle, |n_R\rangle, n_L, n_R=0,1,2,\ldots$) and transmon ($|n_q\rangle, n_q=0,1$), i.e.\ $\{|1\rangle = |000\rangle$, $|2\rangle = |100\rangle$, $|3\rangle = |010\rangle$,
$|4\rangle = |001\rangle$, $|5\rangle = |110\rangle$, $|6\rangle = |101\rangle$,
$|7\rangle = |200\rangle$\}.  As for the reservoir parameters, we consider a fixed temperature profile of $T_L= 330$\ mK and $T_R=100$\ mK as in \cite{ronzani2018tunable} with typical couplings rates $\eta_\alpha$ and Debye frequencies $\omega_{D, \alpha}$ that fit experimental findings, see also the discussion in Sec.~\ref{sec:experiment}.

\subsection{Perturbative treatment versus HEOM }
\label{subsec:mer}

\begin{figure}[htbp]
\centering
\includegraphics[width=8cm]{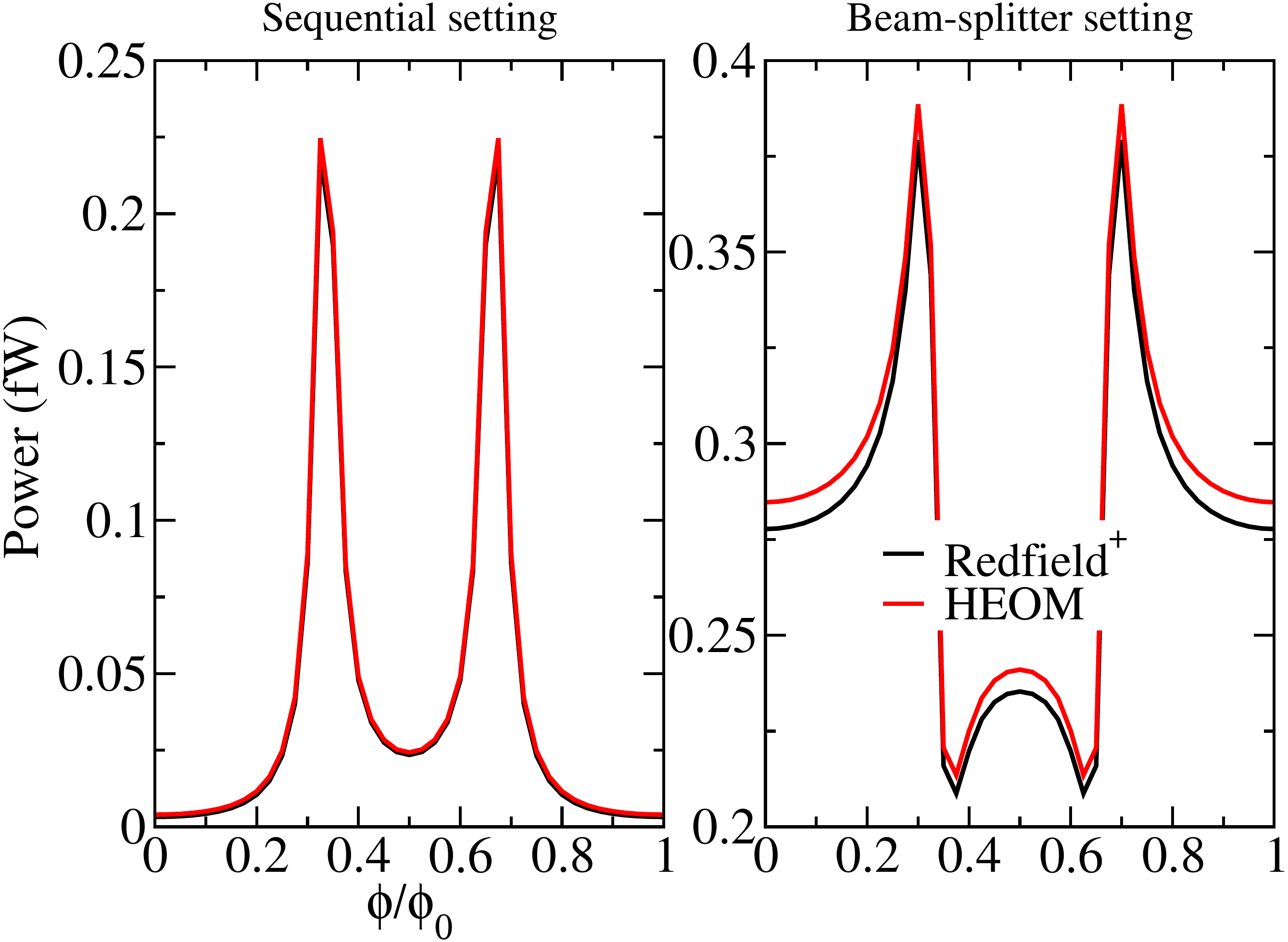}
\caption{Heat power versus magnetic flux: Data from HEOM (red) and Redfield-plus  master equation as in Eq. (\ref{Eq:heom2rd}).
Parameters are $E_{j0}/{2\pi} = 40$ $\rm{GHz}$, $E_{c}/{2\pi} = 0.15$
$\rm{GHz}$, $d = 0.45$, $g_L = g_R = 0.55$ $\rm{GHz}$, $\tilde{g} = -0.55$ $\rm{GHz}$,
$\rm{T}_L = 330$ $\rm{mK}$, $\rm{T}_R = 100$ $\rm{mK}$,
$\hbar\omega_L = \hbar\omega_R = 33.3$ $\rm{GHz}$,
$\eta_L = \eta_R = 0.03$ and
$\omega_{D,L} = \omega_{D,R} = 60$ $\rm{GHz}$.}
\label{fig:fig4}
\end{figure}

\begin{figure}[htbp]
\centering
\includegraphics[width=8cm]{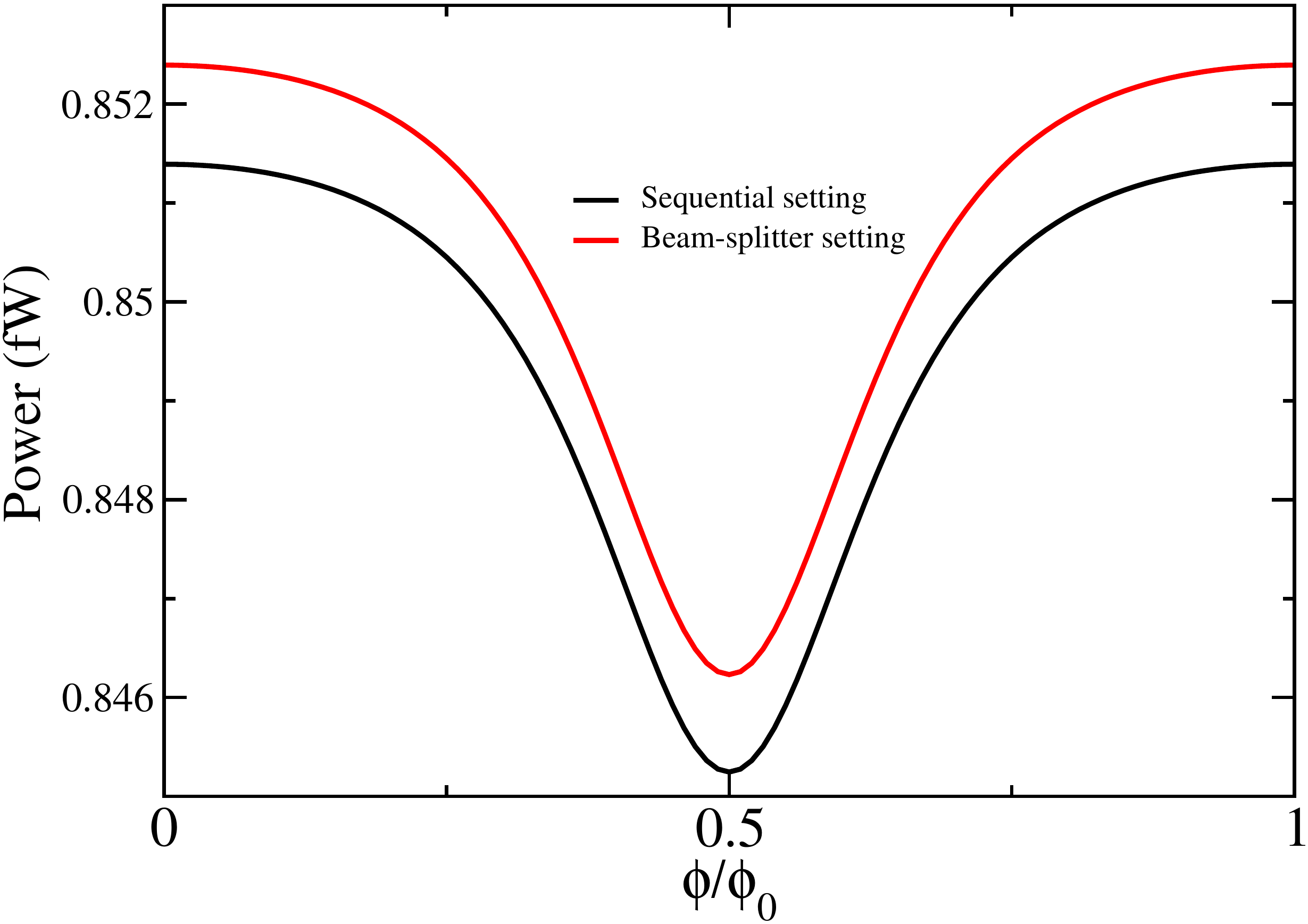}
\caption{Photonic heat power calculated with Fermi's Golden Rule (FGR). Parameters are the same as in Fig. (\ref{fig:fig4}).}
\label{fig:fig5}
\end{figure}

In Figs. \ref{fig:fig4} and \ref{fig:fig5} data are shown  for three types of treatments, namely,  Redfield-plus according to Eq.  (\ref{Eq:heom2rd}) , a conventional Fermi golden rule (FGR) approach, see Appendix A, and the exact HEOM. Note that for the sequential setting we here use the original formulation Eq.~(\ref{Eq:shsys}) with Debye-type thermal reservoirs and an oscillator-qubit-oscillator system part.

In agreement with our expectation from Sec.~\ref{subsec:sec-bsp}, the heat power versus the flux exhibits two peaks together with an almost vanishing value away from it for the sequential setting.  The beam-splitter case leads in addition to a local maximum around $\phi/\phi_0=0.5$ induced by the direct oscillator-oscillator coupling. In fact, in this dispersive regime, the effective coupling $|\tilde{g}_{\rm eff}|$ achieves a maximum at $\phi/\phi_0=0.5$ and decreases away from it due to an increasing $\langle\sigma_z\rangle/\Delta$.

Apparently, in the sequential situation Redfield-plus is in excellent agreement with exact data, while stronger deviations appear for the beam-splitter one, particularly in the dispersive regime. In the latter case, eigenfunctions of the bare system, relevant for heat transfer, are strongly delocalized and may induce reservoir-reservoir correlations that are absent in a perturbative approach.
In contrast, a description in terms of a conventional master equation predicts a substantially larger heat power with quite a different modulation, see Fig.~\ref{fig:fig5}. In both settings, the resonance peaks
are absent. One has to keep in mind that the heat power results from a steady state driven by a thermal gradient, a situation for which the applicability of conventional master equations is not obvious. Further, the Debye frequency extracted to match experimental data (see Sec.~\ref{sec:experiment}) may not exceed the relevant system frequencies sufficiently as required treatments based on Markov approximations.

\begin{figure}[htbp]
\centering
\includegraphics[width=8cm]{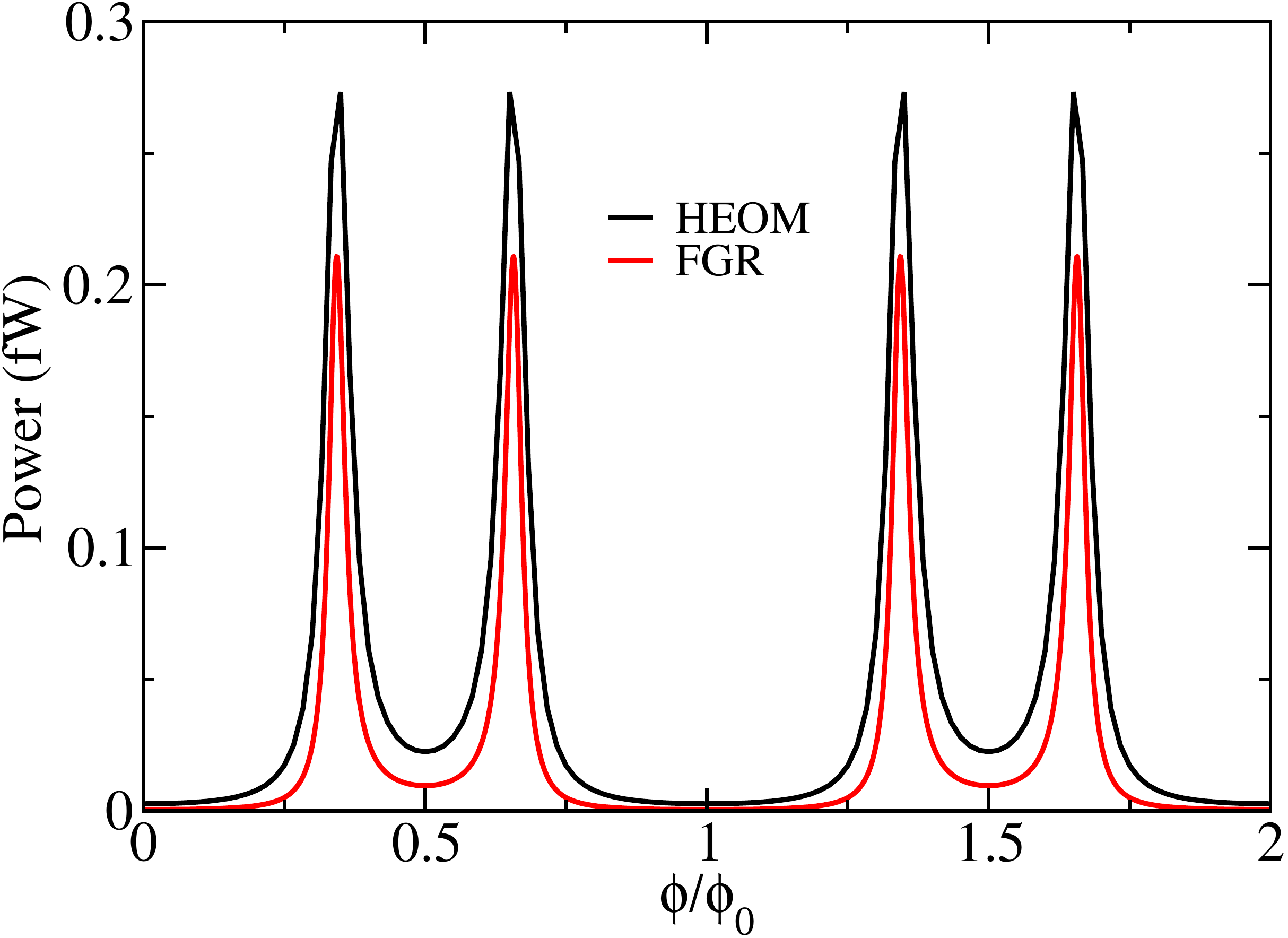}
\caption{In sequential setting, photonic heat transport calculated with HEOM and FGR based on the effective formulation of Eq. (\ref{Eq:effecth}).
Parameters are $E_{j0}/{2\pi} = 40$ $\rm{GHz}$, $E_{c}/{2\pi} = 0.15$
$\rm{GHz}$, $d = 0.45$, $g_L = g_R = 0.55$ $\rm{GHz}$,
$\rm{T}_L = 330$ $\rm{mK}$, $\rm{T}_R = 100$ $\rm{mK}$,
$\hbar\omega_L = \hbar\omega_R = 33.3$ $\rm{GHz}$,
$\eta_L = \eta_R = 1.5$ $\rm{GHz}$.}
\label{fig:fig6}
\end{figure}

For stronger coupling between the reservoirs and the neighboring oscillators,
the condition of weak system-bath interaction $\eta_{\alpha}\ll g_{\alpha},\tilde{g}$ breaks down. As discussed above, this implies that energy level splittings of bare system eigenstates do no longer exceed the level broadening due to the system-bath interaction. In a sequential setting, one way to proceed is then to consider the oscillators as part of the respective reservoir which leads to the spin-boson formulation around (\ref{Eq:effecth}) with a structured spectral density. Within a master equation approach this modelling is only justified as long as a time scale separation exists between the decay of bath correlations [cf.~(\ref{Eq:bwct})] and the overall relaxation of the reduced system. This implies a sufficiently weak oscillator-qubit coupling ($\kappa_\alpha\omega_q \ll \eta_\alpha/\omega_q\ll 1$).  In simple generalization of the golden rule expression for ohmic spectral distributions \cite{segal05a}, one then finds for the heat current through the two level system
\begin{equation}\label{Eq:segal}
 I(\omega_q)
 = \frac{\omega_q J_{{\rm eff},L}(\omega_q)J_{{\rm eff},R}(\omega_q) [ n_L(\omega_q)-n_R(\omega_q)]}
  {J_{{\rm eff},L}(\omega_q)[1+2n_L(\omega_q)] + J_{{\rm eff},R}(\omega_q)[1+2n_R(\omega_q)]} \;\; .
\end{equation}
with the effective spectral distribution $J_{{\rm eff},\alpha}(\omega)$ of the $\alpha $th bath as in Eq. (\ref{Eq:jeff}). Fig. \ref{fig:fig6} shows corresponding results in comparison to data from the HEOM.
We observe that the description based on (\ref{Eq:segal}) captures the flux modulation in the quantum heat valve quite accurately but underestimates
the value of the heat power. At transmon-oscillator resonances $\omega_q\approx \omega_\alpha$, the effective spectral distribution reduces to $J_{\rm eff}(\omega_q)\approx \kappa_\alpha \omega_\alpha^3/\eta_\alpha$ and thus overestimates the heat transfer towards small values of $\eta_\alpha$;  this in turn limits the applicability of (\ref{Eq:segal}).

\subsection{Quantum heat valve: Steady state dynamics}
\label{subsec:phv}
In this and the following sections, the operation of the beam-splitter setting as quantum heat valves is explored in more detail based on numerical HEOM simulations.
\begin{figure}[htbp]
\centering
\includegraphics[width=8cm]{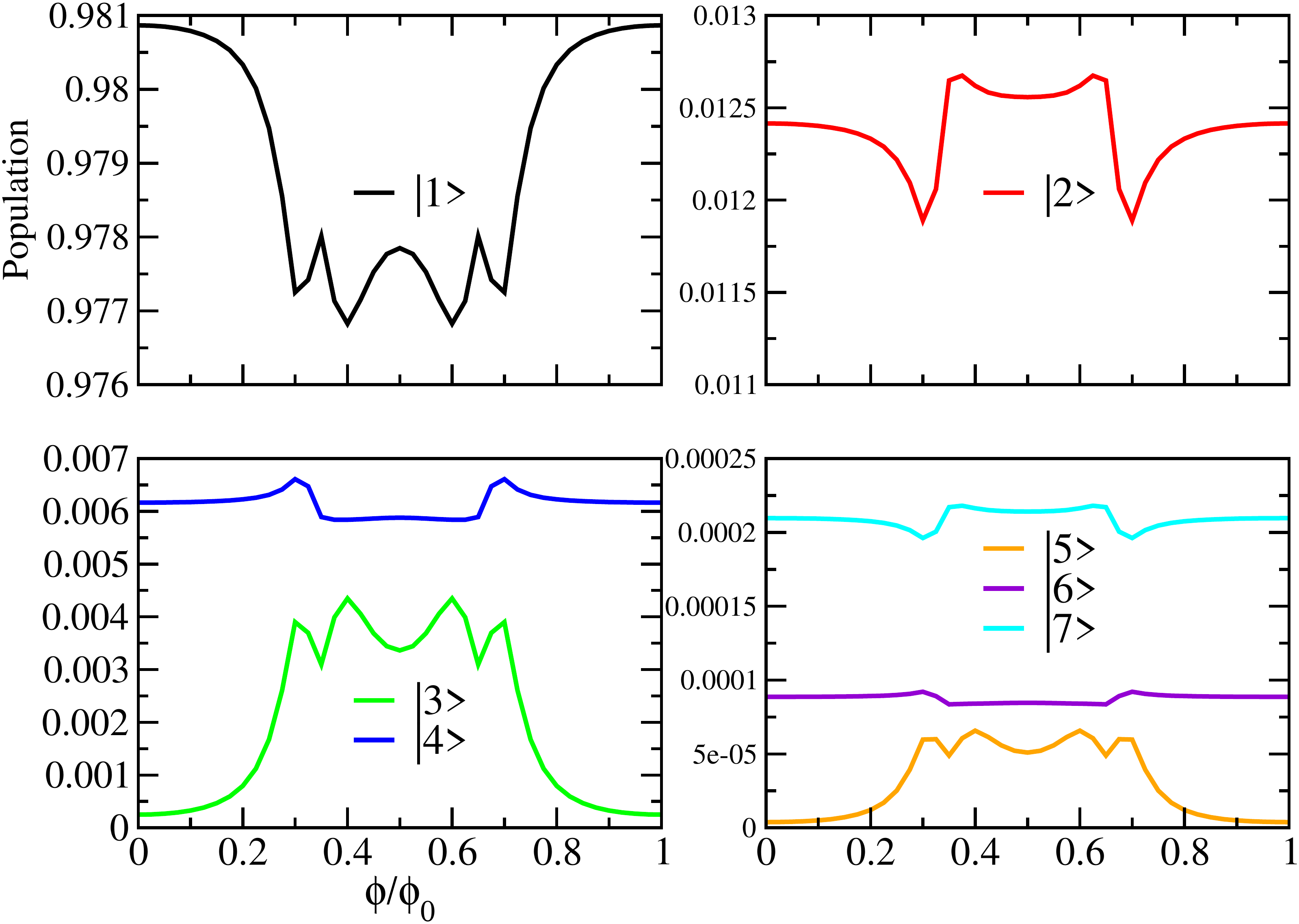}
\caption{Steady state populations versus magnetic flux in a quantum heat valve.
Simulation parameters are same as in Fig. \ref{fig:fig4}.}
\label{fig:fig7}
\end{figure}
Deeper insight can be obtained from the analysis of steady state populations in terms of local states $k\rangle$ as defined above Sec.~\ref{subsec:mer}, see Fig.~\ref{fig:fig7}. Roughly, one can again distinguish between resonant and dispersive regimes when the external flux is tuned. Ground state populations prevail outside the range around and between the resonances. They approach minima at the resonances and smooth local maxima at $\phi/\phi_0=0.5$. The latter directly lead to  local maxima in the heat power for finite oscillator-oscillator coupling. In accordance with the temperature gradient $T_L>T_R$, state $|2\rangle$ ($n_L=1, n_q=0, n_R=0$) is much stronger populated than $|3\rangle, |4\rangle$ for which $n_L=0$; around and between the resonances it mirrors the behavior of the state $|1\rangle$. Double excited states are even less populated which in turn justifies the Hilbert space reduction.

\begin{figure}[htbp]
\centering
\includegraphics[width=8cm]{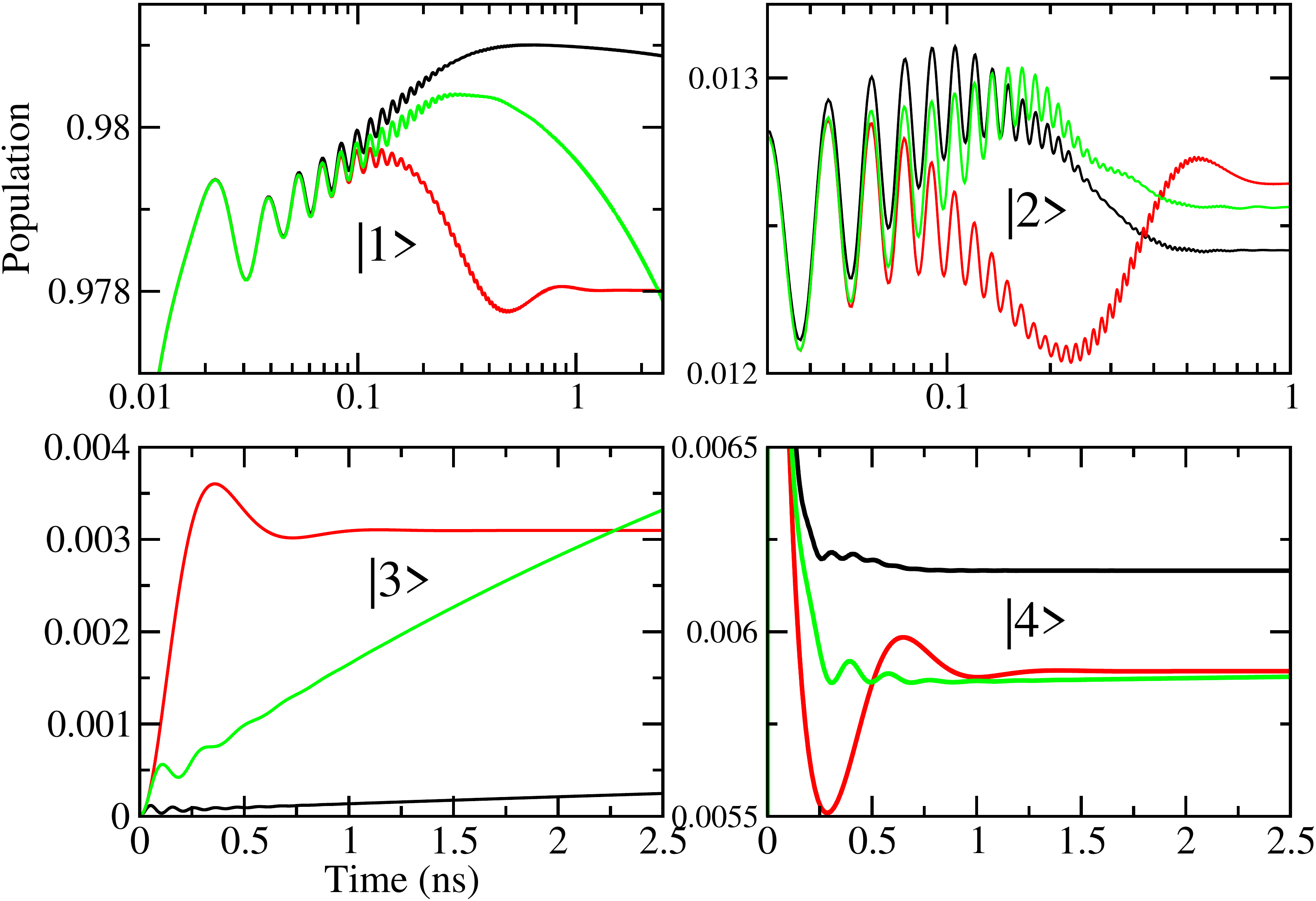}
\caption{Population dynamics towards steady state in the regimes of dispersive and resonant energy transfer.
Black ($\phi/\phi_0 = 0$), red ($\phi/\phi_0 = 0.35$) and blue ($\phi/\phi_0 = 0.5$) curves
 correspond to regimes (a), (b) and (c) in
Fig. \ref{fig:fig3}. Other parameters are as in Fig. \ref{fig:fig4}.}
\label{fig:fig8}
\end{figure}

\begin{figure}[htbp]
\centering
\includegraphics[width=8cm]{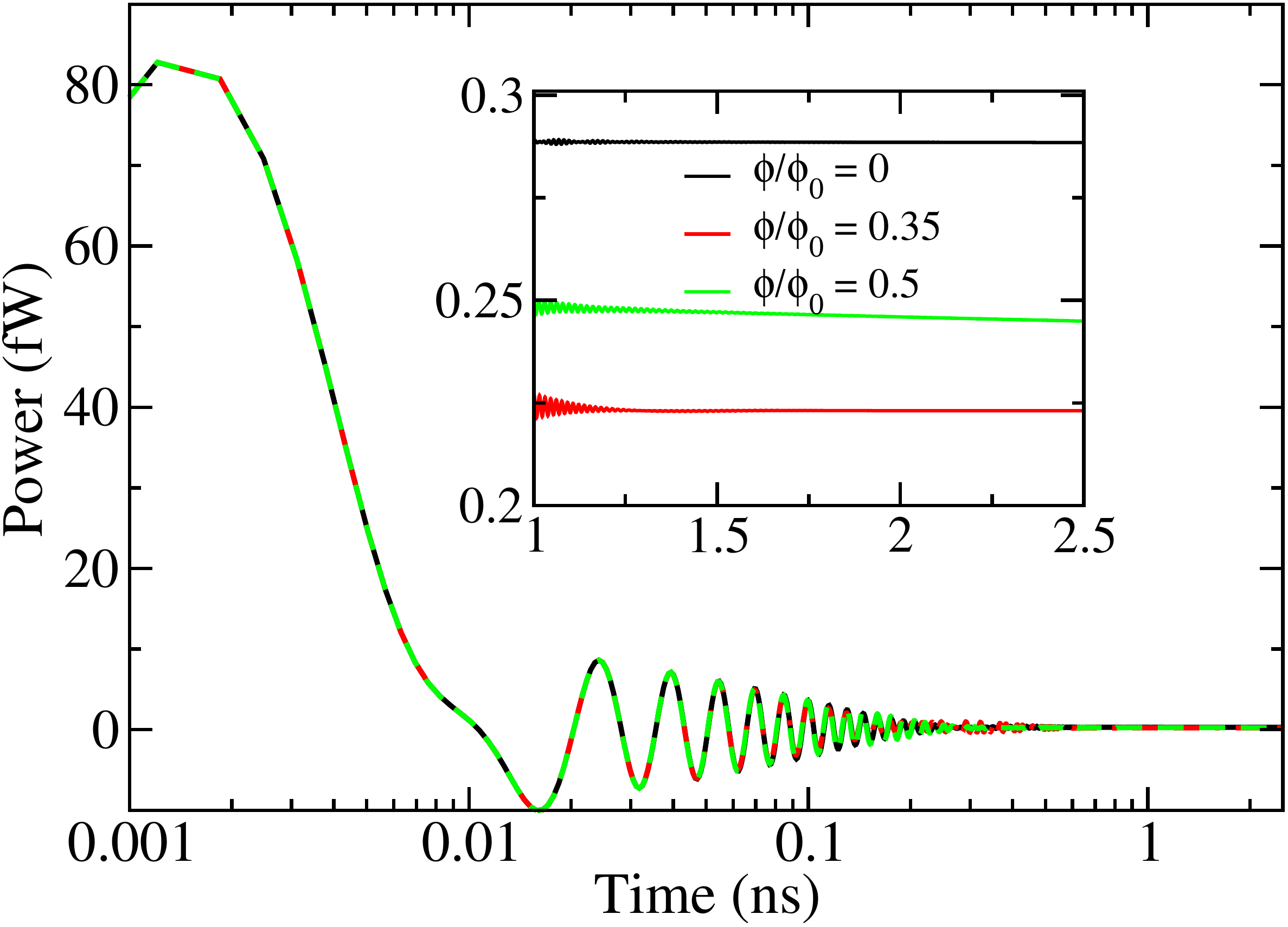}
\caption{Heat current dynamics in the regimes of dispersive and resonant energy transfer.
Black ($\phi/\phi_0 = 0$), red ($\phi/\phi_0 = 0.35$) and blue ($\phi/\phi_0 = 0.5$)
curves correspond to regimes (a), (b) and (c) in Fig. \ref{fig:fig3}. The inset displays the long-time behavior. Simulation parameters are the same as in Fig. \ref{fig:fig4}.}
\label{fig:fig9}
\end{figure}

The non-equilibrium dynamics of both, the populations and the heat power, are displayed in Figs.~\ref{fig:fig8}, \ref{fig:fig9} in the dispersive and resonant regimes, respectively. Here, we oberserve the overall behavior that steady states are approached on similar time scales in both regimes. Interestingly, substantial transient oscillations occur that we associate with coherent quantum dynamics between the relevant eigenstates. When operating the devices on shorter time scales, it may be possible to exploit these coherences for proper control techniques.

\subsection{Quantum heat valve: parameter dependence}
\label{subsec:tpe}
Here we discuss the performance of the experimental device when relevant parameters are tuned, also in order to tailor the design of future set-ups. Relevant quantities include
transmon-qubit parameters  $d$ and $E_{j0}$,
qubit-resonant coupling parameters $g_{\alpha}$, the beam-splitter parameter $\tilde{g}$, and resonator-bath coupling parameters $\eta_{\alpha}$. These parameters are typically fixed during the fabrication process of the circuitry.

We start with the asymmetry parameter $d$ which enters the flux dependence of the effective Josephson coupling energy $E_J(\phi)$ in Eq. (\ref{eq:ejphi}). Note that for any finite value of $d$, the coupling energy stays always finite.
\begin{figure}[htbp]
\centering
\includegraphics[width=8cm]{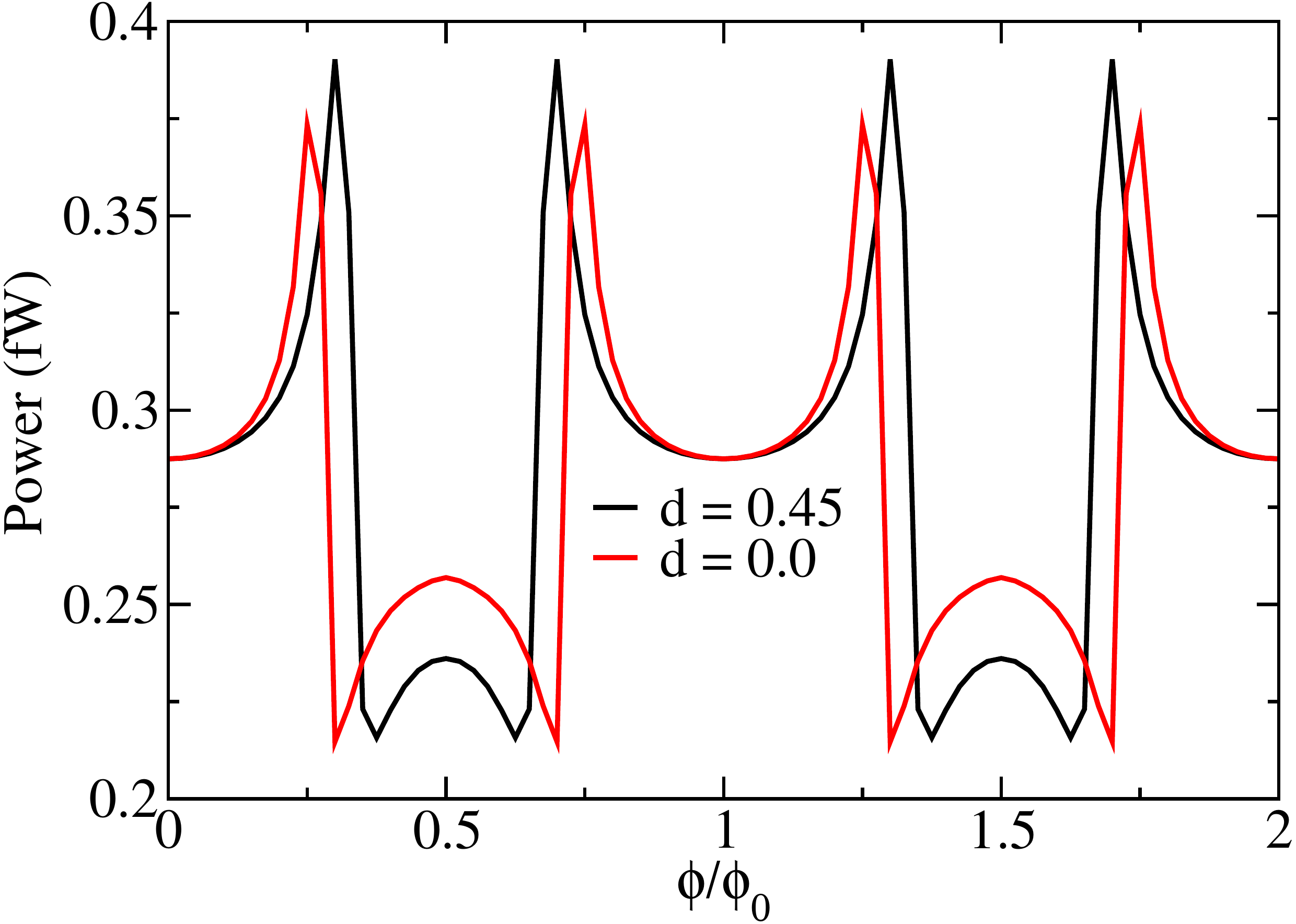}
\caption{Dependence of the heat power versus magnetic flux through a quantum heat valve on the asymmetry parameter $d$ between the two Josephson junctions constituting the tunable transmon. Other parameters are as in Fig. \ref{fig:fig4}.}
\label{fig:fig10}
\end{figure}
Fig. \ref{fig:fig10} displays the impact on the modulation of the photonic heat valve.
A decreasing value of $d$ has basically three consequences: The domain between the resonances grows, the relative peak heights decrease, and the local maxima in the dispersive regime at $\phi/\phi_0=0.5$ tend to be more pronounced. This behavior can be traced back to an increased de-tuning $\Delta$ with decreasing asymmetry. This in turn makes 
 the effective coupling $g_{\rm eff}$ to grow, thus leading to a larger heat transition rate. Note that the parameter $d$ is in general difficult to fine-tune in chemical etching and difficult to measure precisely. Theory provides means to extract it from experimental data.

\begin{figure}[htbp]
\centering
\includegraphics[width=8cm]{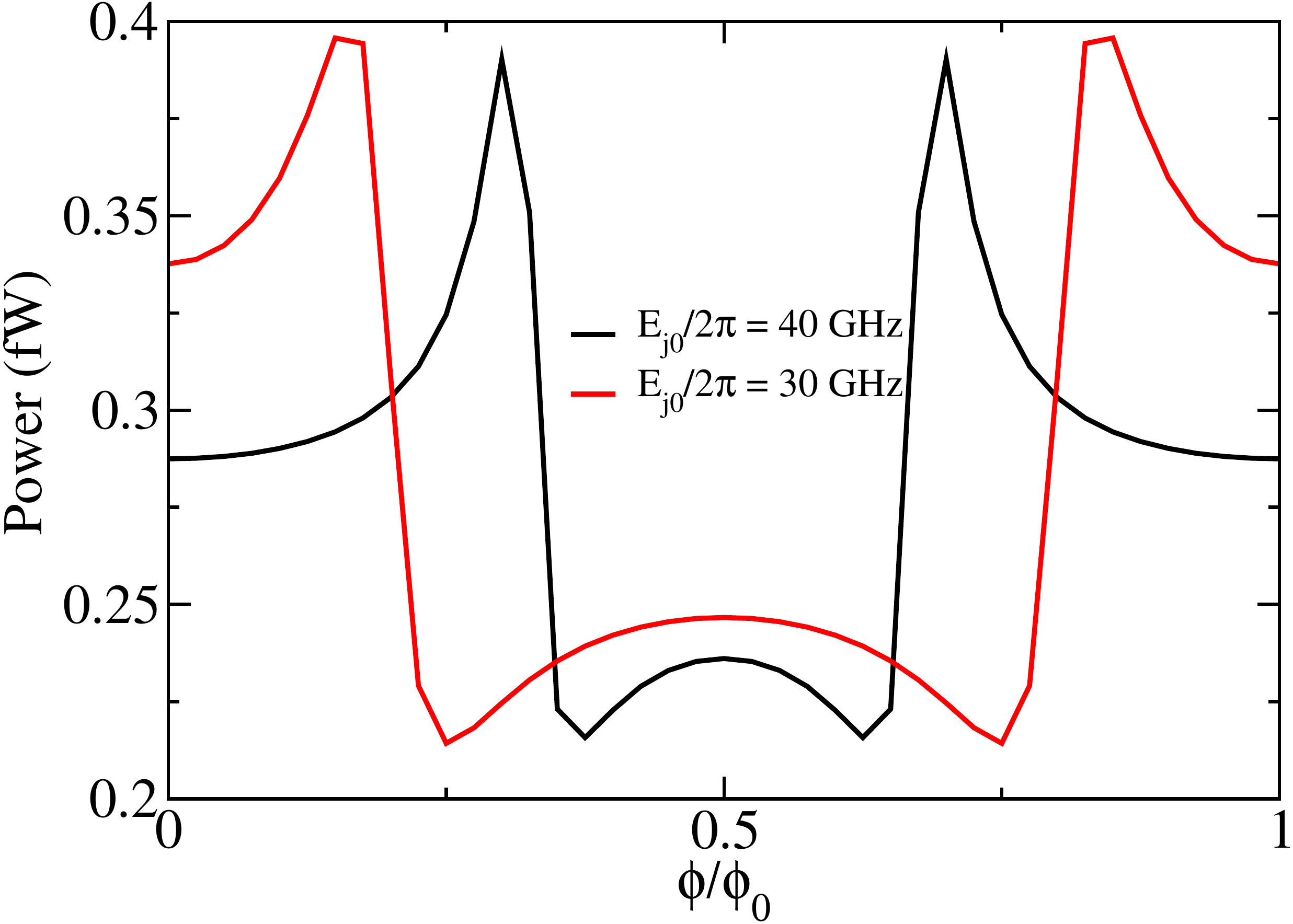}
\caption{Dependence of the heat power through a quantum valve on the  Josephson coupling energy $E_{j0}$ for $d=0$. Other parameters are as in
 Fig. \ref{fig:fig4}.}
\label{fig:fig11}
\end{figure}
Next, in Fig.~\ref{fig:fig11}, the dependence on the bare Josephon energy $E_{j0}$ is shown.  The dominant effect is a broadening of the domain between the resonances, while the relative peak heights and the local maxima in the dispersive regime are less sensitive. A smaller value for $E_{J, 0}$ makes the resonances to appear further away from the symmetry point $\phi/\phi_0=0.5$ since for the given parameters $\omega_L=\omega_R$ the maximal de-tuning $\Delta$ decreases.

\begin{figure}[htbp]
\centering
\includegraphics[width=8cm]{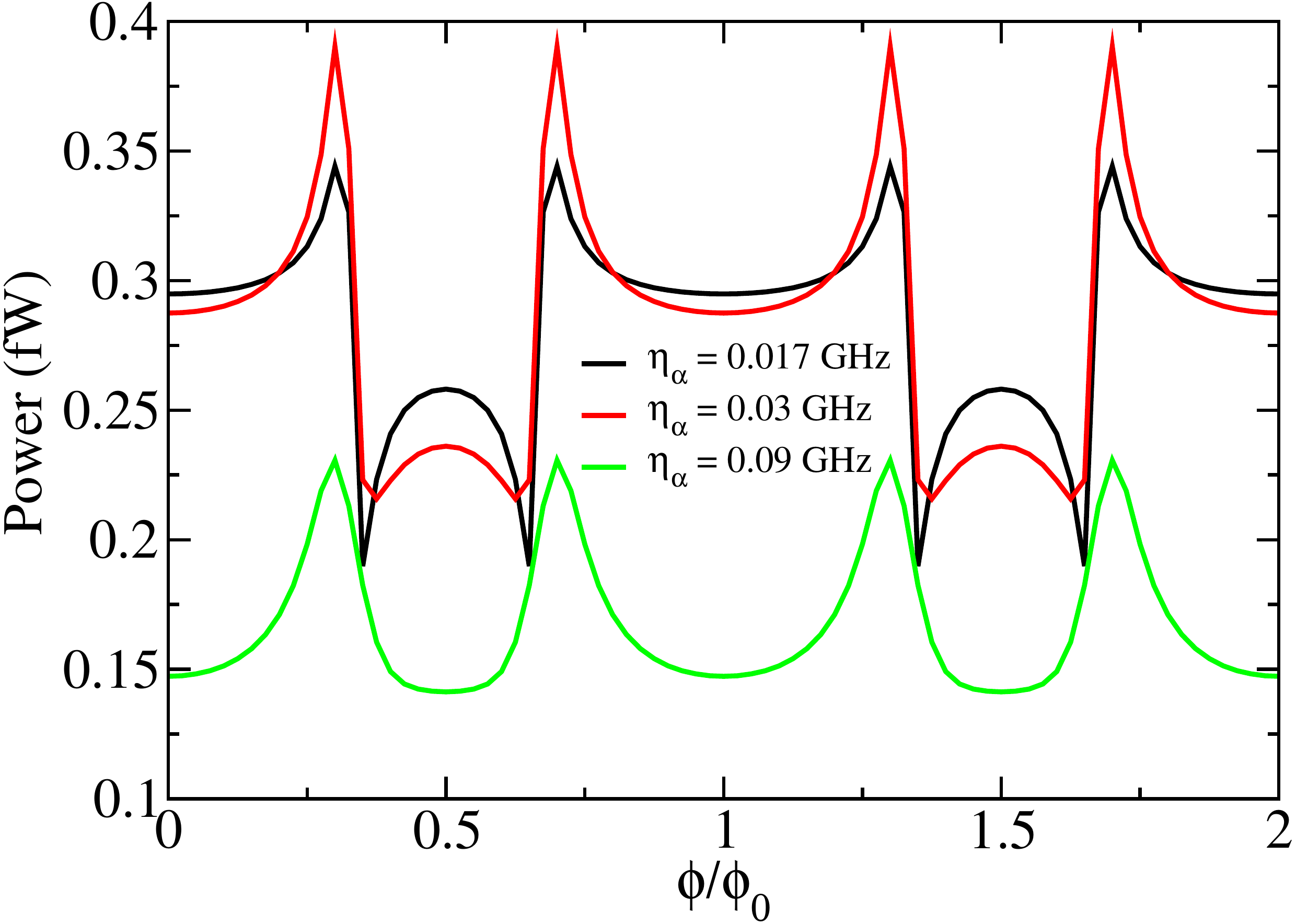}
\caption{Dependence of the heat power through a quantum heat valve on the thermal coupling $\eta_{\alpha}$.
Other parameters are the same as in Fig. \ref{fig:fig4}.}
\label{fig:fig12}
\end{figure}
Very interesting is the dependence on the oscillator-reservoir coupling $\eta\equiv \eta_L=\eta_R$ displayed in Fig.~\ref{fig:fig12}. First of all, we observe that the overall modulation with tunable magnetic flux is smeared out towards stronger thermal contact. This particularly applies to the local maxima in the dispersive regime. A stronger coupling has the tendency to localize eigenstates, while heat transfer occurs most efficiently through delocalized  eigenstates. The same process applies to  avoided crossings between oscillator-transmon levels and in turn reduces the heights of the resonance peaks.

The competition between dissipation induced localization versus resonance-induced delocalization has been seen in the experiment \cite{ronzani2018tunable} and discussed theoretically also in
 \cite{velizhanin2008heat,hanggi1990reaction,pollak2002classical}.
In fact, it leads to a  turnover behavior of the heat power with respect to the system–bath coupling strength that can be rationalized in the following way: When the system-bath coupling strength is tuned from weak to
intermediate, the coupling plays an effective role to accelerate
photons to tunnel between system and bath. However, for stronger system-bath
coupling, the bath-induced friction that dissipates  energy away from the system dominates which implies a slower photon tunnelling rate. In between these regimes,  maxima of the thermal conductance appear \cite{velizhanin2008heat}.

\begin{figure}[htbp]
\centering
\includegraphics[width=8cm]{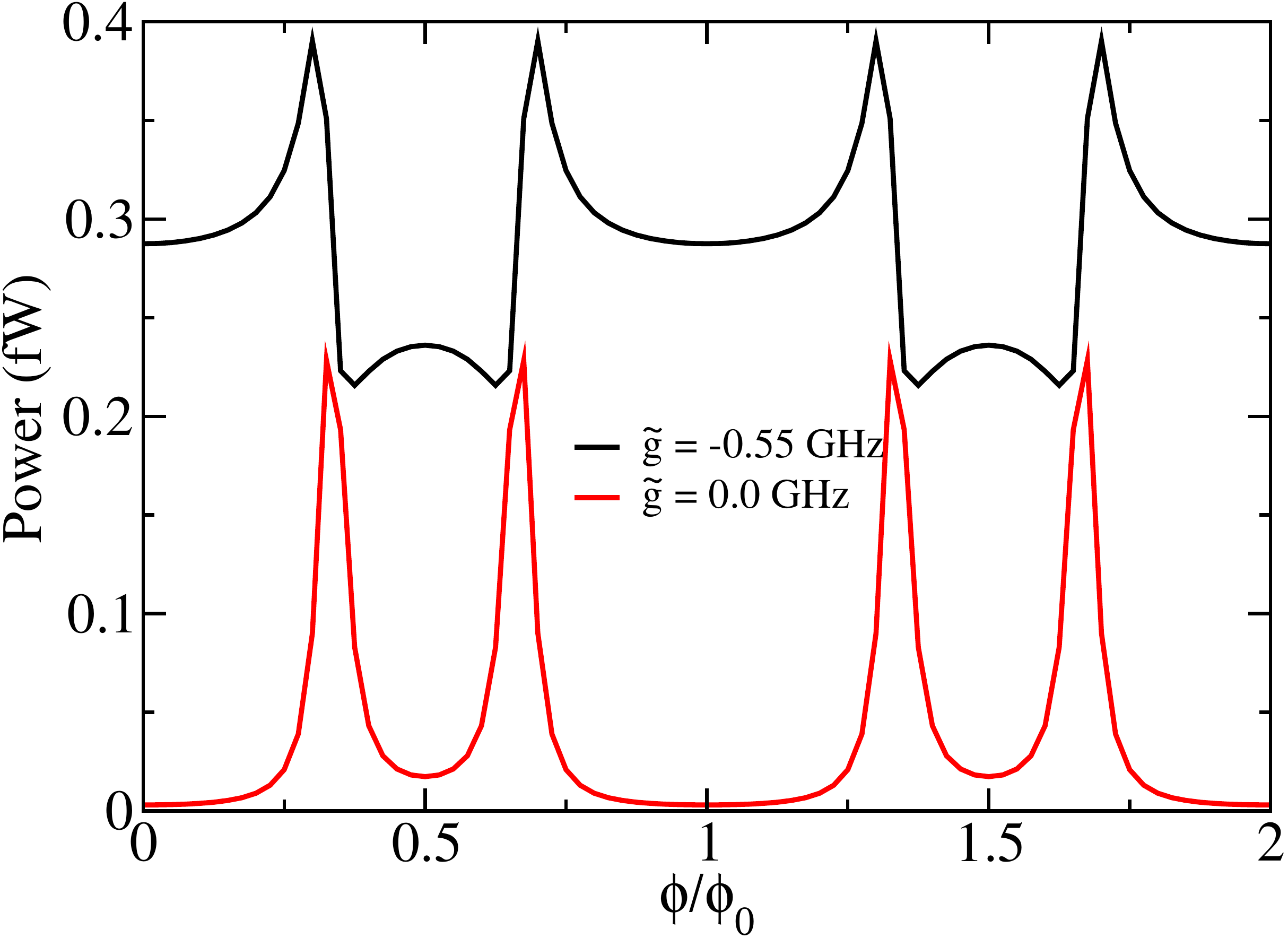}
\caption{Dependence of the heat power through a quantum heat valve on the direct oscillator-oscillator coupling $\tilde{g}$.
Other parameters are as in Fig. \ref{fig:fig4}.}
\label{fig:fig13}
\end{figure}
\begin{figure}[htbp]
\centering
\includegraphics[width=8cm]{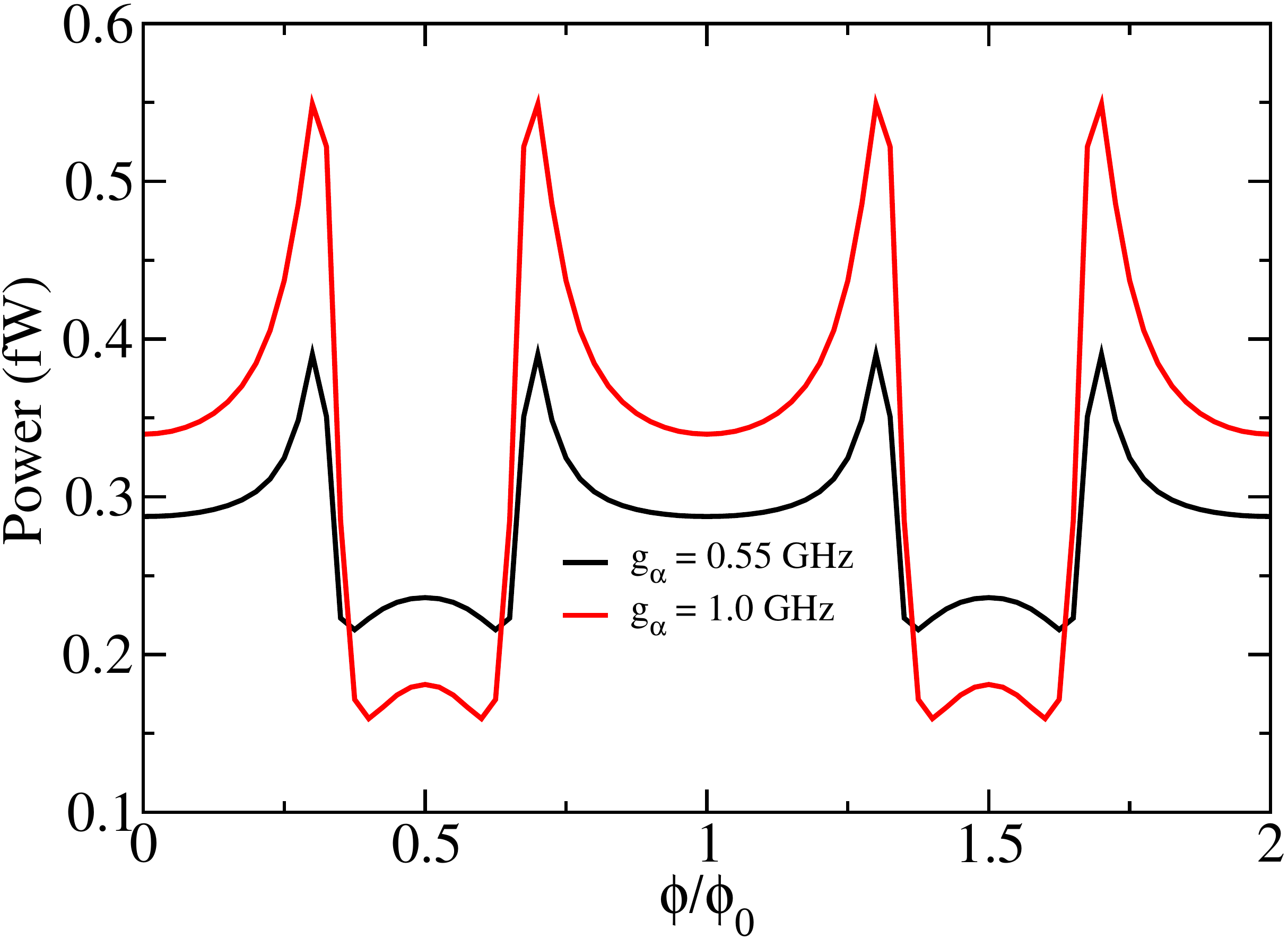}
\caption{Dependence of the heat power through a quantum heat valve on the qubit-resonators coupling $g_{\alpha}$.
Other parameters are as in Fig. \ref{fig:fig4}.}
\label{fig:fig14}
\end{figure}

The dependence on the intra-system coupling parameters $\tilde{g}$ (direct oscillator channel) and $g$ (oscillator-transmon channel) can be seen in
Figs.\ref{fig:fig13} and \ref{fig:fig14}, respectively.
Of course, for a vanishing oscillator-oscillator coupling (sequential setting) the local maxima at $\phi/\phi_0=0.5$ are not present and heat can be transferred only through the transmon. The latter increases for stronger $g$ and implies larger relative peak heights in the heat power.

\subsection{From quantum heat valve to quantum rectifier}
\label{subsec:asse}

Heat rectification appears when the net heat flow between forward and backward transfer for reversed temperature gradient is finite. This requires a symmetry breaking in the device architecture, for example by putting $g_L\neq g_R$ or, experimentally easier to realize, $\omega_L\neq \omega_R$. In the past, heat rectification has been discussed theoretically in the classical regime, see for example \cite{Terraneo2002controlling,li2004thermal}. In the quantum domain simple models like the spin-boson model \cite{segal05a} and harmonic systems \cite{riera2019dynamically} have been treated. A non-perturbative approach applicable to a broad class of set-ups has been formulated only recently in \cite{motz2018rectification} followed by experimental realizations \cite{senior2020heat}. Here, based on the cQED platform, we address heat rectification by introducing an asymmetric system structure with $\omega_L \neq \omega_R$, where $\omega_L$ is considered as tunable.
\begin{figure}[htbp]
\centering
\includegraphics[width=8cm]{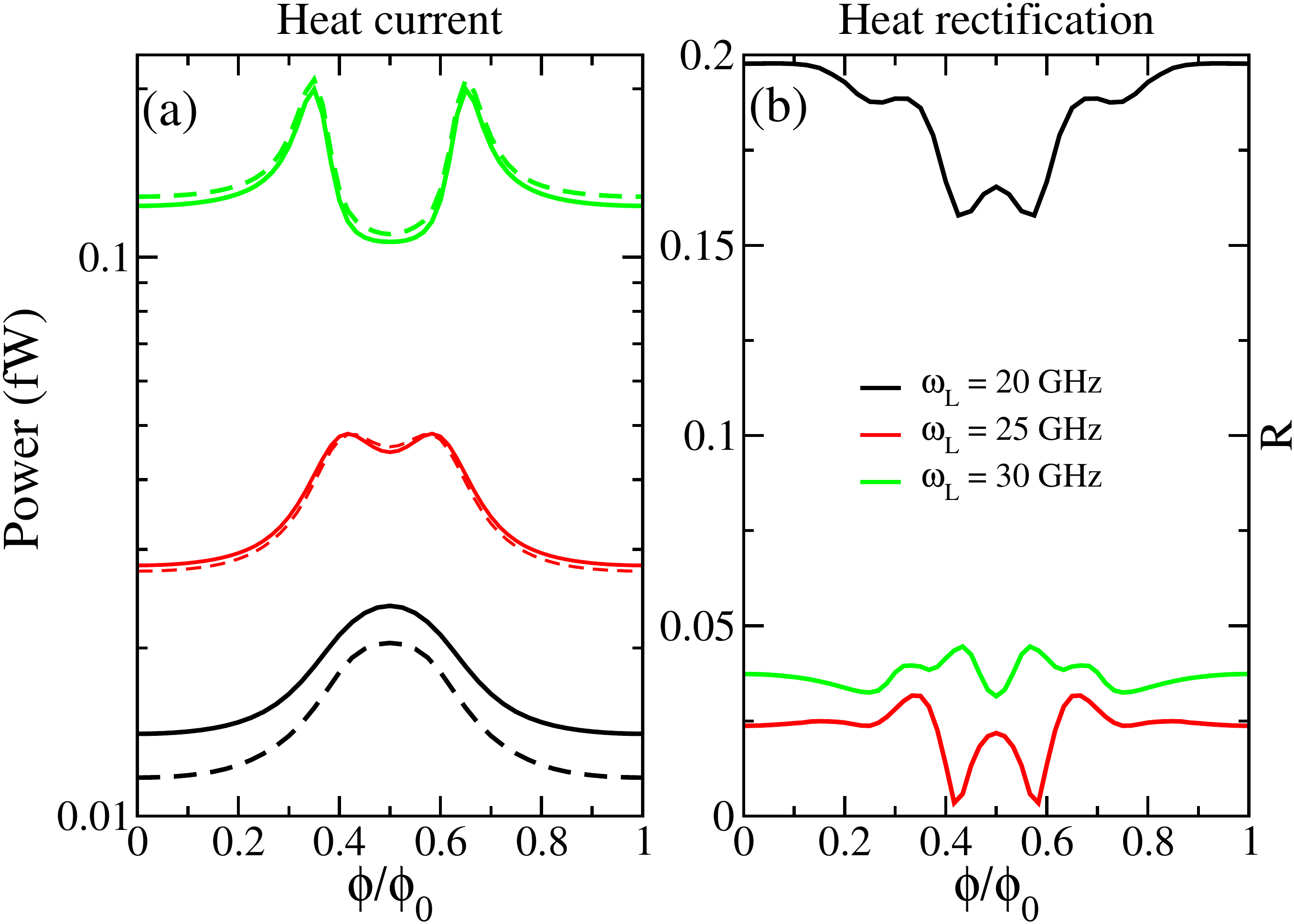}
\caption{Rectification of the photonic heat power for an asymmetric system architecture with $\omega_L\neq \omega_R$.
In panel (a), solid curves corresponds to forward photonic heat power, dashed curves to backward photonic heat power. Panel (b) shows corresponding heat rectification in terms of the coefficient (\ref{Eq:rectification}).
In the simulations $\omega_L$ changes from $20$ GHz to $30$ GHz, while other
parameters are kept constant as in Fig. \ref{fig:fig4}.}
\label{fig:fig15}
\end{figure}

Panel \ref{fig:fig15}(a) depicts the asymmetry in the forward and backward heat power which grows with increasing de-tuning $\omega_L\neq \omega_R$. Notably, with increasing asymmetry the peak-valley feature, characteristic for the symmetric situation, disappears and is replaced by a single central hump at $\phi/\phi_0=0.5$.
Further,  the symmetry-breaking now leads to resonances between qubit and individual oscillators which thus appear at different values of the external flux or may not be present at all for large detunings.

To quantify the rectification, we introduce the rectification coefficient
\begin{equation}\label{Eq:rectification}
  \mathcal{R} = \frac{|P_f - P_b|}{|P_b|} \;\;
\end{equation}
as the weighted net transfer between forward ($P_f$) and backward ($P_b$) heat power.
Corresponding results are shown in Fig.~\ref{fig:fig15}(b).
As expected, a stronger asymmetry induces a larger overall rectification. Interesting is the modulation of the coefficient $\mathcal{R} $ with the magnetic flux: For very weak asymmetry ($\omega_L=30$ GHz) shallow peaks appear near  transmon-oscillator resonances followed by a central dip. With growing asymmetry ($\omega_L=25$ GHz) two more pronounced maxima develop together with two pronounced dips and a local maximum at $\phi/\phi_0=0.5$. For strong asymmetry ($\omega_L=20$ GHz) this central maximum shrinks as well as the overall maxima. However, in all situations we see that a tunable magnetic flux allows to control the rectification: From almost 100\% for intermediate asymmetry to 25\% for strong asymmetry. This may offer new options to regulate heat transfer in cQED devices. We also note that the dependence on the asymmetry is non-monotonous: In the regime of smaller detunings, an increase in asymmetry first leads to a decrease in the rectification performance; for stronger asymmetry rectification is always enhanced.

\section{Photonic heat transport: Comparison with experimental data}
\label{sec:experiment}

In the previous sections we studied to which extent the heat transfer through a generic superconducting set-up  can be quantitatively described by approximate treatments in comparison to non-perturbative HEOM simulations. For this purpose, the thermal reservoirs were assumed to carry a generic spectral distribution of Debye-type.

Here, we lay focus on an actual experimental situation described in \cite{ronzani2018tunable}. There, a two level system, implemented in form of a transmon qubit with tunable frequency, is capacitively embedded between two superconducting transmission lines
each terminated by a mesoscopic normal-metal reservoir. Effectively, this leads to the model depicted in  Fig.~\ref{fig:fig1}, however, as seen from the harmonic oscillators, with a
Lorentzian-type of spectral distributions with maxima around their bare frequencies $\omega_L=\omega_R$.

However, in many cases less detailed information about the nature of thermal reservoirs is known so that one may assume only a class of
spectral distributions, rather than an explicit distribution.
Then, the following question arises: How strongly does heat transfer in the quantum regime depend on the specific form of the spectral distribution as long as a substantial portion of the distribution covers the relevant 'bandwidth' of eigenenergies? To be more specific in the present case: Is it possible to quantitatively describe experimental data of \cite{ronzani2018tunable} with different forms of spectral distributions parametrized by physically meaningful parameters?
\begin{figure}[htbp]
\centering
\includegraphics[width=8cm]{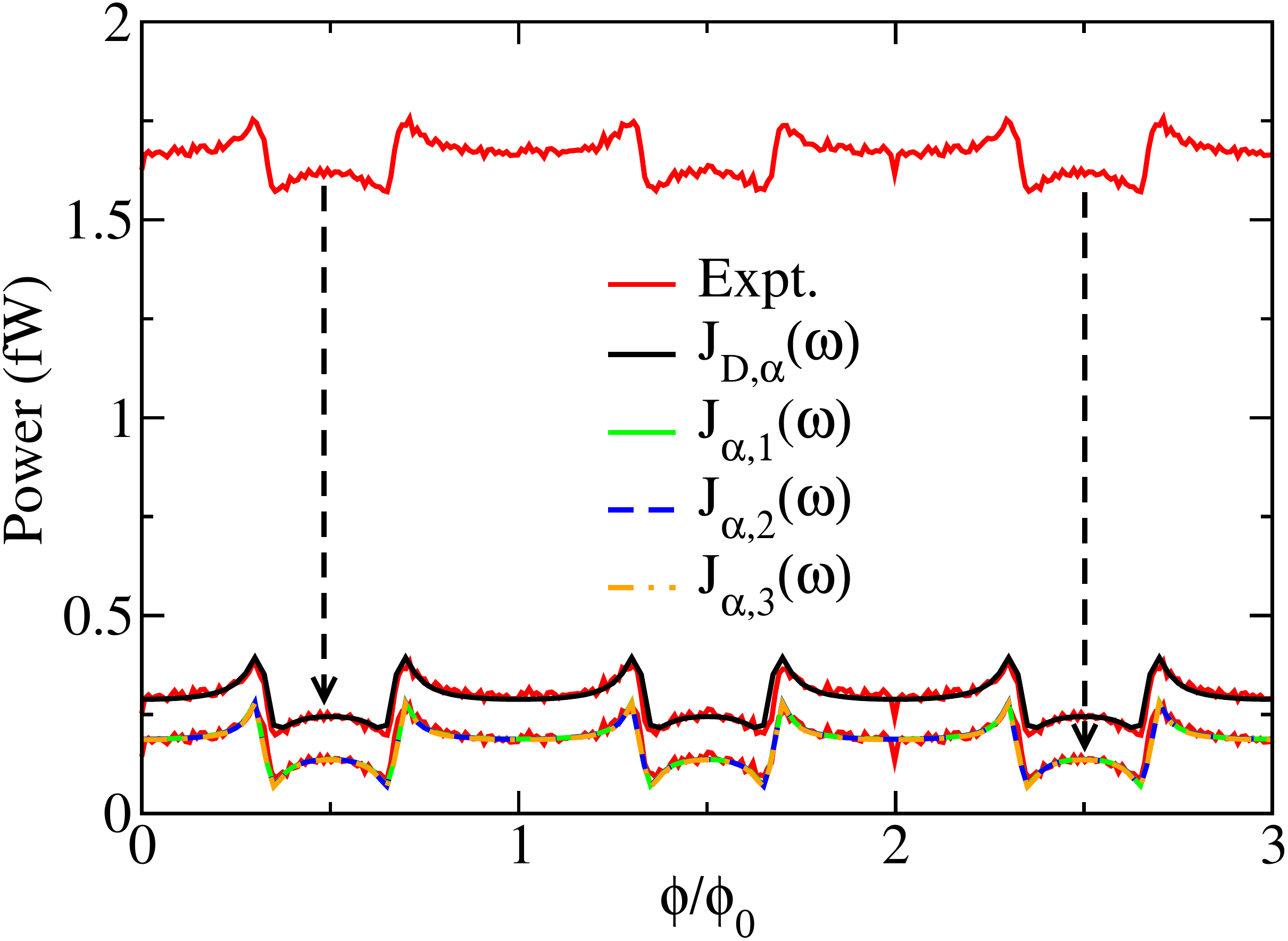}
\caption{Heat power  in a quantum heat valve with temperatures $\rm{T}_L = 330$ $\rm{mK}$, $\rm{T}_R = 100$ $\rm{mK}$, resonator frequencies $\omega_L = \omega_R = 33.3$ $\rm{GHz}$, and
$E_{c}/{2\pi} = 0.15$ $\rm{GHz}$
simulated with two types of reservoir spectral densities in the HEOM. For the Debye spectral density other parameters are:
$E_{j0}/{2\pi} = 40$ $\rm{GHz}$,
$d = 0.45$, $g_L = g_R = 0.55$ $\rm{GHz}$, $\tilde{g} = -0.55$ $\rm{GHz}$, $\eta_L = \eta_R = 0.03$, and
$\omega_{D,L} = \omega_{D,R} = 60$ $\rm{GHz}$;
for the Lorentz  spectral densities $J_{\alpha, n}$:
$E_{j0}/{2\pi} = 34$ $\rm{GHz}$,
$d = 0.58$, $g_L = g_R = 0.35$ $\rm{GHz}$,
$\tilde{g} = -0.25$ $\rm{GHz}$, $Q_L = Q_R = 40$.}
\label{fig:fig16}
\end{figure}

For this purpose, we focus on Debye-type as in (\ref{eq:debye}) and on Lorentz-type distributions, respectively, i.e.,
\begin{equation}\label{Eq:lorentz}
    J_{\alpha, n}(\omega) = \frac{1}{Q_{\alpha}^3}\frac{\omega_{\alpha}^{5-n}\ \omega^n}
    {(\omega^2-\omega_{\alpha}^2)^2+\omega_{\alpha}^2\omega^2/
    Q_{\alpha}^2} \;\;,
\end{equation}
with a quality factor $Q_{\alpha}=\omega_{\alpha}/\eta_{\alpha}$ and for $n=1,2,3$.
Note that $J_{\alpha, n}\propto \omega^n$ in the low frequency sector $\omega\ll \omega_\alpha$ and $J_{\alpha, n}\propto 1/\omega^{4-n}$ for high frequencies; in particular, at low frequencies $J_{\alpha, 1}$ describes ohmic, while $J_{\alpha, 2/3}$ super-ohmic behavior, and at resonance $J_{\alpha, n}(\omega_\alpha)= \omega_\alpha/Q_\alpha$.
The corresponding correlation functions read
\begin{equation}\label{Eq:corr-lorentz}
\begin{split}
  C_{\alpha, n}(t) =& \frac{1}{\pi} \int_{0}^{+\infty}d\omega J_{\alpha, n}(\omega)
 [\coth\frac{\beta_{\alpha}\omega}{2}\cos\omega t - i\sin\omega t] \\
 =& \sum_{\sigma=\pm}  \frac{\omega_{\alpha}^{4-n}(\xi_{\alpha}-i\sigma\eta_{\alpha}/2)^{n-1}}
 {4Q_{\alpha}^2\xi_{\alpha}}\\
 &\times\left\{\coth[\frac{\beta_{\alpha}}{2}
  (\xi_{\alpha}-i\sigma \frac{{\eta}_{\alpha}}{2})] + \sigma\right\}\\
&\times  \exp\left[-(\frac{{\eta}_{\alpha}}{2}+i \sigma\xi_{\alpha})t\right] \\
&  +\frac{2\omega_{\alpha}^{5-n}}{\beta_{\alpha}Q_{\alpha}^3}
\sum_{k=1}^{\infty} \frac{ (-i)^{n+1}\nu_k^n}{(\omega_{\alpha}^2 + \nu_k^2)^2 -{\eta}_{\alpha}^2\nu_k^2} e^{-\nu_k t} \;\;.
\end{split}
\end{equation}

\begin{figure}[htbp]
\centering
\includegraphics[width=8cm]{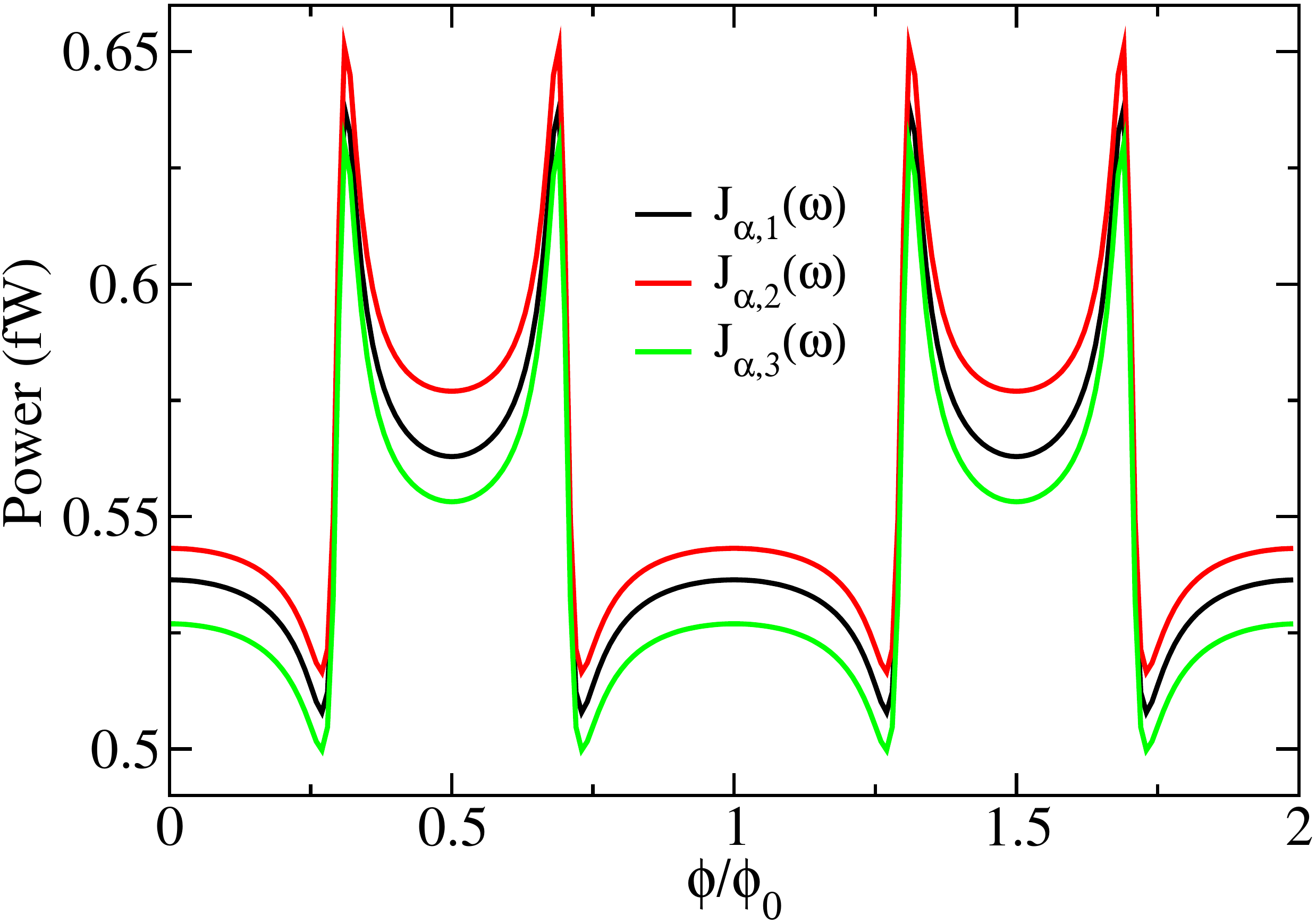}
\caption{Photonic heat transport through a quantum heat valve studied by FGR with Lorentz-class spectral density; parameters are as in Fig. \ref{fig:fig16}.}
\label{fig:fig17}
\end{figure}

Fig. \ref{fig:fig16} shows a comparison between theoretical predictions and experimental data for the heat power from \cite{ronzani2018tunable} with properly adjusted parameters. Important to mention here is the fact that in the actual experimental set-up in addition to the {\em flux sensitive photonic} channel there is a  {\em flux independent phononic} channel for
 energy transfer.  The corresponding latter contribution was not measured in the experiment and is here considered as an overall constant subtracted from the experimental data (see arrows).

The first main point we observe is that the experimental findings can be very accurately captured by all four types of spectral distributions. In fact, predictions based on the Lorentz forms with $J_{\alpha, n}$ all coincide using identical sets of parameters which implies that only the behavior around the resonance frequencies matter, while the low and high frequency portions of the respective distributions have only very minor impact. Indeed, the most relevant heat carrying eigenstates exhibit frequencies in a range around $\omega_L, \omega_R$ as seen in Figs.~\ref{fig:fig2}, \ref{fig:fig7}. Note that even the Debye distribution provides a quantitative agreement with a somewhat smaller contribution from the (flux insensitive) phononic channel. From our simulations we predict the magnitude of this portion to the net heat transfer to be about 1.4 fW, substantially bigger than the modulation amplitude of about 0.2 fW.

Using the same parameter set, we show in Fig.~\ref{fig:fig17} results based on a golden rule treatment (see App.~A). The main difference to the HEOM data appears around $\phi/\phi_0= k/2$ ($k$ integer), where local minima (maxima) occur instead of maxima (minima). In addition, the golden rule predictions based on the Lorentz forms $J_{\alpha,n}$ do not coincide using identical sets of parameters,
particularly in the dispersive regime in contrast to HEOM predictions.

\section{Conclusions}
\label{sec:conclusions}

This paper provides benchmark results for quantum heat transfer through a generic set-up consisting of a resonator-transmon-resonator channel terminated by thermal reservoirs. Theoretical predictions have been obtained within the formally exact HEOM simulation technique and also in comparison with the performance of perturbative treatments.

With respect to the first major question placed in the Introduction about the accuracy of the latter ones, we conclude that while they may provide a qualitative description in some aspects, they certainly fail to give a quantitative understanding of experimental results. Remarkably, this is even true in domains of parameter space, where conventional approaches naively are assumed to work. With respect to the second question about the parameter dependence, one observes that (i) the structure of the internal couplings play a decisive role, (ii) the precise structure of the thermal reservoirs, i.e.\ the spectral distributions, does much less matter as long as in the frequency domain their main portions cover the relevant eigenstate channels, (iii) too strong thermal couplings wash out the modulation features and reduce its amplitude. HEOM (and other non-perturbative approaches) allow to fix circuit parameters which are not sufficiently known.
The beam splitter setting provides means to fine-tune heat transfer by adjusting, may be even in a tunable manner, the resonator-transmon versus the resonator-resonator coupling.

\section*{Acknowledgment}
M. X. thanks Prof. Qiang Shi for the discussions about HEOM and his group computing resources. We thank Ronzani Alberto and Jukka P. Pekola for discussions and providing their experimental data. We also thank Yaming Yan, M. Wiedmann and G. Kurizki for fruitful discussions. This work has been supported by IQST and the German Science Foundation (DFG) under AN336/12-1 (For2724).

\section*{Appendix A: Fermi Golden Rule (FGR)}
We can derive the Fermi Golden Rule description (FGR) directly from Eq. (\ref{Eq:redfield}). In the Schr\"odinger representation, one then arrives at the expressions
\begin{equation}
\begin{split}
  \frac{d}{dt}\rho_s(t)
  &= -i\mathcal{L}_s\rho_s(t)-\int_0^{+\infty}d\tau
 C(\tau)\;[\hat{q},(e^{-i\mathcal{L}_s\tau}\hat{q})\rho_s(t)] \\
  & + C(-t) \;[\rho_s(t)(e^{-i\mathcal{L}_s\tau}\hat{q}),\hat{q}]\;\;,
\end{split}
\end{equation}

If we just consider diagonal elements $P_j(t) = \{\rho_s(t)\}_{jj}$, these equations reduce to
\begin{equation}
   \frac{d}{dt}P_j(t) = -\mathcal{R}_{jj}P_j(t)-\sum_{k\neq j} \mathcal{R}_{jk} P_k(t) \;\;,
\end{equation}
with
\begin{equation}\label{}
\begin{split}
   \mathcal{R}_{jk} &= \int_{0}^{+\infty} d\tau
   C(\tau) \langle j|[\hat{q},\; (e^{-i\mathcal{L}_s\tau}\hat{q})\; |k\rangle\langle k|]|j\rangle \\
   &+
   C(-\tau)\langle j|[|k\rangle\langle k|\;(e^{-i\mathcal{L}_s\tau}\hat{q}),\;\hat{q}] |j\rangle  \\
&=\sum_{r}q_{jr}q_{rk}\delta_{kj}\int_{0}^{+\infty} d\tau C(\tau)e^{-i\omega_{rk}\tau}\\
&-q_{jk}q_{kj}\int_{0}^{+\infty} d\tau C(\tau)e^{-i\omega_{jk}\tau} \\
&+\sum_{r}q_{kr}q_{rj}\delta_{kj}\int_{0}^{+\infty} d\tau C(-\tau)e^{-i\omega_{kr}\tau}\\
&-q_{jk}q_{kj}\int_{0}^{+\infty} d\tau C(-\tau)e^{-i\omega_{kj}\tau} \\
&= \sum_{r}q_{jr}q_{rk}\delta_{kj}\int_{-\infty}^{+\infty} d\tau C(\tau)e^{-i\omega_{rk}\tau}\\
&-q_{jk}q_{kj}\int_{-\infty}^{+\infty} d\tau C(\tau)e^{-i\omega_{jk}\tau}\;\;,
\end{split}
\end{equation}
where $H_s|j\rangle = \hbar\omega_j|j\rangle$, $\omega_{jk} = \omega_j - \omega_k$ and $q_{jk}=\langle j|\hat{q}|k\rangle$.
Applying the Fluctuation-Dissipation theorem,
\begin{subequations}
\begin{equation}\label{}
  C(t) = \frac{1}{\pi}\int_{-\infty}^{+\infty}d\omega
\frac{J(\omega)}{1-e^{-\beta \omega}}e^{-i\omega t}\;\;;
\end{equation}
\begin{equation}\label{}
 \frac{J(\omega)}{1-e^{-\beta\omega}} = \frac{1}{2}\int_{-\infty}^{+\infty}dt \; C(t)e^{i\omega t} \;\;,
\end{equation}
\end{subequations}
we find that
\begin{subequations}\label{}
\begin{equation}
  \mathcal{R}_{jj} = \sum_{r} |\langle j|\hat{q}|r\rangle|^2 \frac{2J(\omega_{jr})}{1-e^{-\beta\omega_{jr}}}\;\;;
\end{equation}
\begin{equation}\label{}
  \mathcal{R}_{jk} =
  - |\langle j|\hat{q}|k\rangle|^2 \frac{2J(\omega_{kj})}{1-e^{-\beta\omega_{kj}}}\;\; {\rm for}\;\; k\neq j\;\;.
\end{equation}
\end{subequations}
Finally, the Pauli master equation is obtained
\begin{equation}\label{Eq:pme}
\begin{split}
  \frac{d}{dt}P_j(t)
=&  - \mathcal{R}_{jj}P_j(t)
  -\sum_{k\neq j}\mathcal{R}_{jk}P_k(t)  \\
=&  - \sum_{k} |\langle j|\hat{q}|k\rangle|^2
\frac{2J(\omega_{jk})}{1-e^{-\beta\omega_{jk}}}P_j(t)\\
&+   \sum_{k\neq j} |\langle j|\hat{q}|k\rangle|^2
\frac{2J(\omega_{kj})}{1-e^{-\beta\omega_{kj}}}P_k(t) \\
=&  2\sum_{k\neq j} |\langle j|\hat{q}|k\rangle|^2
[ J(\omega_{jk})n_{\beta}(\omega_{jk})P_k(t)\\
& - J(\omega_{kj})n_{\beta}(\omega_{kj})P_j(t)] \\
=& \sum_{k\neq j}
\left[\Gamma_{jk}P_k(t) - \Gamma_{kj}P_j(t) \right]  \;\;,
\end{split}
\end{equation}
with FGR rates $\Gamma_{jk}$. In the situation of two thermal reservoirs and in non-equilibrium steady-state $P_k(t)\to P_k^{(\infty)}$, the heat current flowing between system and the $\alpha$th bath is given by
\begin{equation}\label{Eq:fgr}
    I_{\alpha} = \sum_{j,k} P_k^{(\infty)} \; \omega_{jk} \;
    \Gamma_{jk;\alpha} \;\;.
\end{equation}

\section*{References}
\bibliography{quantum}

\end{document}